\documentclass{article}

\usepackage{arxiv}

\usepackage[utf8]{inputenc} 
\usepackage[T1]{fontenc}    
\usepackage{hyperref}       
\usepackage{url}            
\usepackage{booktabs}       
\usepackage{amsfonts}       
\usepackage{nicefrac}       
\usepackage{microtype}      
\usepackage{lipsum}
\usepackage{graphicx}
\graphicspath{ {./images/} }

\usepackage{amsmath}
\usepackage{natbib}

\usepackage{siunitx}
\usepackage{moreverb}
\usepackage{mathtools, cuted}
\usepackage{rotating}

\title{Causal dependencies and Shannon entropy budget - Analysis of a reduced order atmospheric model}

\author{
Stéphane Vannitsem \\ Royal Meteorological Institute of Belgium, Avenue Circulaire, 3, 1180 Brussels \\ 
\texttt{Stephane.Vannitsem@meteo.be} \\
\And
Carlos A. Pires \\ Instituto Dom Luiz, Faculdade de Ciências, Universidade de Lisboa, Lisbon, Portugal \\
 \texttt{clpires@ciencias.ulisboa.pt} \\
\And 
 David Docquier \\ Royal Meteorological Institute of Belgium, Avenue Circulaire, 3, 1180 Brussels \\
 \texttt{David.Docquier@meteo.be} \\
}

\begin{document}
\maketitle

\begin{abstract}
The information entropy budget and the rate of information transfer between variables is studied in the context of a nonlinear reduced-order atmospheric model. The key ingredients of the dynamics are present in this model, namely the baroclinic and barotropic instabilities, the instability related to the presence of an orography, the dissipation related to the surface friction, and the large-scale meridional imbalance of energy. For the parameter chosen, the solutions of this system display a chaotic dynamics reminiscent of the large-scale atmospheric dynamics in the extra-tropics. The detailed information entropy budget analysis of this system reveals that the linear rotation terms plays a minor role in the generation of uncertainties as compared to the orography and the surface friction. Additionally, the dominant contribution comes from the nonlinear advection terms, and their decomposition in synergetic (co-variability) and single (impact of each single variable on the target one) components reveals that for some variables the co-variability dominates the information transfer. The estimation of the rate of information transfer based on time series is also discussed, and an extension of the Liang's approach to nonlinear observables, is proposed. 
\end{abstract}


\section{Introduction}

Causal analysis of the occurrence of certain events, in particular extreme events, is the central concern of many research activities in meteorological and climate sciences. For example, one prominent question in the last decades is the causal impact of CO$_2$ on the evolution of the global temperature \citep{Stips2016}, or the attribution of climate change on the occurrence of more intense extreme events \citep{Bellprat2019}. To answer such type of questions, different approaches can be used such as running climate models with and without CO$_2$ forcing, and/or investigating the fingerprint of the evolution of CO$_2$ on actual observations through linear regression \cite[e.g.][]{Allen2003,Chen2024}. Other important questions related for instance to the role of the ocean on the atmospheric evolution and the emergence of low-frequency variability in the atmosphere \cite[e.g.][]{Bellucci2008, VannitsemGhil2017, Paolini2024} and concerning the role of tropical regions in shaping the extra-tropical variability \cite[e.g.][]{Alexanderetal2002, Schemmetal2018, Yeh2018} are also attracting considerable interest. 

In all these questions, one central statistical quantity commonly used to infer causality is the correlation and all its variants (e.g. regression, lagged correlation). This quantity, although a very useful indicator, does not provide a definitive answer on the causal influence from one variable to another. In the last decades, there has been a growing literature on new tools for evaluating the directional dependencies between a wide range of observables. On a theoretical side, the developments include techniques using the improvement of forecasting skill \cite[e.g.][]{Granger1969, Papagiannopoulou2017, Papana2021}, analogs \citep{Sugihara2012}, networks \citep{Runge2018}, or information entropy dynamics \citep{Schreiber2000, Palus2001, Liang2005, Katerina2007, Liang2021, Silini2021}. Some of these approaches have been compared in recent works \citep{Krakovska2018, Runge2019a, Runge2019b, Docquier2024}, which provide useful information on which method to use and for what purpose. All these approaches have also been used in recent years to clarify the directional connections between different climate or meteorological processes: the interaction between the ocean and the atmosphere \citep{Mosedale2006, Tirabassi2015, VannitsemGhil2017, Docquier2023}, the interaction between the land/vegetation and the atmosphere \citep{Papagiannopoulou2017, Hagan2019, Hagan2022, Zhou2024}, the interaction between large scale components of the climate system \citep{Liang2014, Runge2019a, Vannitsem2022, DiCapua2020a, DiCapua2020b, Silini2022, Docquier2024}, the drivers of Arctic sea ice and Antarctic surface mass balance \citep{Vannitsem2019, Docquier2023}, to quote a few.

One particularly important approach has been developed in a series of papers by Liang \citep{Liang2005, Liang2014, Liang2015, Liang2016, Liang2021} based on the development of a theory of the information entropy evolution and the transfer of information between selected variables in the rather general context of deterministic and stochastic dynamical systems. These studies culminated with the derivation of simple relations that could be directly applied to real-world data \citep{Liang2014,Liang2021}, under the assumption that the system under investigation can be described by a linear stochastic system. However, this type of approximation, which is sometimes justified, as illustrated for instance in \cite{Sardeshmukh2015}, can often lead to misinterpretation of interactions. This is particularly true in chaotic dynamical systems for which nonlinearities play a key role in generating the dynamics, such as the \cite{Lorenz1963} model. In such a context, one must therefore turn back to the original equations of the Shannon entropy evolution \citep{Liang2021}, and find ways to evaluate properly the rate of information transfer between observables. An important step in that direction has been recently made in \cite{Pires2024}, in which the rate of information transfer between variables or group of variables can be computed by means of conditional expectations of the system's forcing terms in a fully nonlinear framework. The approach has been validated in the context of several low-order models with or without stochastic forcing. 

Based on these recent advances, the information entropy budget and the transfer of information are explored in this study in a simple reduced-order atmospheric model containing key ingredients of the atmospheric dynamics, namely the baroclinic instability, vorticity transport, rotation, orographic forcing and orographic friction \citep{Charney1980}. This model is a deterministic system for which chaotic solutions are found, provided appropriate parameter values are chosen. A detailed analysis of the information entropy is first performed, clarifying the role of physical processes (e.g. orography, friction, nonlinear advection) on the changes of information entropy. In a second step, the possibility to extract the rate of information transfer and the terms of the information entropy balance from time series is then explored, using the approach designed in \cite{Liang2021}.

In Section \ref{Theory}, the basic equations that will be used in the current analysis are introduced. The atmospheric model is then described in Section \ref{CSmodel}. The different terms of the information entropy balance are discussed in Section \ref{Results}, and the extraction of these terms from time series is discussed in Section \ref{timeseries}. Finally, key conclusions are drawn in Section \ref{conclusion}.

\section{Theory: key equations}
\label{Theory}

\subsection{The entropy balance}
\label{2.1}

A central quantity of information theory is the information entropy, or Shannon entropy, denoted $H$, which is a unique measure of information with the following properties: it takes a maximum value when the probability distribution of the variable is uniform; adding an event with probability 0 does not change its value; and the property of additivity by which $H(AB) = H(A) + H(B|A)$ where $AB$ is the composition of two subsystems $A$ and $B$ and $H(B|A)$ is the conditional entropy \citep{Gray2011, Nicolis2012}. It can be applied to discrete-value systems such as for instance the uncertainty produced by the transition dynamics between weather regimes \cite[e.g.][ and references therein]{Nicolis2012, Vannitsem2023} or for continuous-value systems \cite[e.g.][]{Kleeman2011, Liang2021}.

A detailed description on how to construct an evolution equation for the information entropy is provided in \cite{Liang2016,Liang2021} and \cite{Pires2024}, and we redirect the reader to these papers for the detailed calculations. Here, we will briefly introduce the key quantities that will be used to characterize how information is changing and transferred throughout a dynamical system. 

Let us consider a dynamical system of the form
\begin{equation}
 \frac{d\bf{{X}}}{dt} = {\bf F} ( \bf{X}, \bf{\lambda})
 \label{dynsystem}
\end{equation}
where ${\bf X} = (X_1, ..., X_N)$ with $N$ the number of variables. The system is supposed to be autonomous, and its solutions converge to an ergodic attractor with a stationary probability density. Here, the system is considered as deterministic, although the theory is valid when stochastic components are present \citep{Liang2021,Pires2024}.   

In this context, the rate of change of the Shannon entropy associated to variable $X_i$ is given by
\begin{equation}
\frac{d H_{X_i}}{dt} = \frac{d H_{X_i,self}}{dt} + T(X_{ \neq i} \rightarrow X_i)  \label{eq:entropyrate} 
\end{equation}
where $X_{ \neq i}$ denotes the variables that are complementary to $X_i$. Asymptotically, the tendency of entropy should tend toward 0 on such an ergodic system and a balance is reached between the self-entropy generation (SEG; first term on the right-hand side of Eq.~(\ref{eq:entropyrate})) and the rate of entropy transfer (RET; second term on the right-hand side of Eq.~(\ref{eq:entropyrate})) from the set of complementary variables $X_{ \neq i}$ to $X_i$. The expressions for these two quantities are

\begin{eqnarray}
\frac{d H_{X_i,self}}{dt} & = & E \left[  \frac{\partial F_i}{\partial X_i}  \right]  \label{SEG}\\
T(X_{ \neq i} \rightarrow X_i) & = & E \left[  F_i \frac{\partial log(\rho_{\neq i|i})}{\partial X_i}  \right] \nonumber \\ & = & E \left[  F_{i, \neq i} \frac{\partial log(\rho_{\neq i|i})}{\partial X_i}  \right] 
\label{decompositionentropy}
\end{eqnarray}
where $F_{i, \neq i}$ is the part of the deterministic term $F_i$ which contains all the terms influencing $X_i$ that are not exclusively dependent on $X_i$ and $\rho_{\neq i|i}$ is the conditional probability density function of all the variables $X_k, k \neq i$ given $X_i$. $E[.]$ denotes the expectation over all variables.

The RET can be further expressed in terms of the conditional expectations of the variables assuming that $F_{i, \neq i}$ is a sum of separable functions (such as multivariate polynomial functions) \citep{Pires2024}:

\begin{equation}
T(X_{ \neq i} \rightarrow X_i) =  \sum_{l} E_i \left[  f_{1,l} (X_i) \frac{\partial E_{k \neq i} (f_{2,l}(X_{ k \neq i})| X_i)}{\partial X_i}  \right]
\label{totalret}
\end{equation}
where $F_{i, \neq i}$ is written as $ \sum_{l} f_{1,l} (X_i) f_{2,l}(X_{k \neq i})$. For purely linear functions in $X_{k \neq i}$, we have $ f_{1,l} (X_i) = 1$. $E_i[.]$ and $E_{k \neq i}[.]$ denote the expectation over variables $X_i$ and $X_{k \neq i}$, respectively.

The RET from the complementary variables is an important quantity, which may be decomposed in different contributions related to the different physical processes acting on the target variable as we will see in our application. Relating the different physical processes to the transfer of information allows for better understanding the role of these processes on the propagation of information uncertainties. 

In the case of the RET from one variable to another, referred as `single RET' hereafter, a similar procedure can be followed, but now the decomposition is done on $F_{i, j}$ which is the part of $F(i)$ that depends on $X_j$, as follows:

\begin{equation}
T(X_{j} \rightarrow X_i) =  \sum_{l} E_i \left[  E_{X_{k \neq j}}(f_{1,l} (X_{k \neq j})|X_i) \frac{\partial E_{j} (f_{2,l}(X_{j})| X_i)}{\partial X_i}  \right]
\label{singleret}
\end{equation}
where $F_{i, j} =  \sum_{l} f_{1,l} (X_ {k \neq j}) f_{2,l}(X_j) $.

The synergetic contribution is the difference between the total RET from all variables $T(X_{ \neq i} \rightarrow X_i) $ and the sum of all single RETs:
\begin{equation}
T_{Syn}(X_{ \neq i} \rightarrow X_i) = T(X_{\neq i} \rightarrow X_i) - \sum_{j \neq i} T(X_{j} \rightarrow X_i)
\label{Tsyn}
\end{equation}
This quantity contains all contributions that are related to the coherent dynamics, i.e. co-variability, between the variables through nonlinear terms. It is equal to 0 for linear equations. 

Note that the total RET is generally positive as it balances the negative SEG coming from dissipation in a chaotic regime. However, when decomposing the total RET in different contributions, some terms could contribute negatively and others positively to the evolution of the uncertainty. In other words, some terms are expected to control (or reduce) the uncertainty when negative, while others would increase the uncertainty when positive. The sign has therefore a clear meaning in the evolution of the entropy.

\subsection{The simplified linear stochastic approximation}

A simplification of the approach was proposed by \cite{Liang2014} to allow for the direct application on time series. It consists in assuming that the underlying model at the origin of the dynamics is a linear stochastic system that can be fitted to the data. This approach was first proposed to compute dependencies for bivariate series \citep{Liang2014}, and has subsequently been extended to multivariate cases \citep{Liang2016,Liang2021}. Both have been extensively used in various applications \citep{Docquier2022,Hagan2022,Vannitsem2022}. 

Under the assumption of a linear model with additive noise, the maximum likelihood estimate of the RET is given by \citep{Liang2021}:
\begin{equation}
\label{eq-lintransfer}
 T_{X_j \rightarrow X_i} = \frac{1}{\mbox{det} \bf{C}} \cdot \sum_{k=1}^N{\Delta_{jk}C_{k,di}} \cdot \frac{C_{ij}}{C_{ii}},
\end{equation}
where \textbf{C} is the covariance matrix, $\Delta_{jk}$ are the cofactors of \textbf{C} ($\Delta_{jk} = (-1)^{j+k} M_{jk}$, where $M_{jk}$ are the minors), $C_{k,di}$ is the sample covariance between all $X_k$ and the Euler forward difference approximation of d$X_i$/d$t$, $C_{ij}$ is the sample covariance between $X_i$ and $X_j$, and $C_{ii}$ is the sample variance of $X_i$.

To assess the importance of the relative importance of cause-effect relationships, a normalization of the RET can be performed. This new quantity is referred as $\tau_{X_j \rightarrow X_i}$ from variable $X_j$ to variable $X_i$ and reads as \citet{Liang2015,Liang2021}
\begin{equation}
\label{eq-lintransferrelative}
 \tau_{X_j \rightarrow X_i} = \frac{T_{X_j \rightarrow X_i}}{Z_i},
\end{equation}
where $Z_i$ is the normalization:
\begin{equation}
\label{eq7}
 Z_i = \sum_{k=1}^N{\left|T_{X_k \rightarrow X_i}\right|} + \left|\frac{dH_i^{noise}}{dt}\right|,
\end{equation}
where the first term on the right-hand side represents the information flowing from all the $X_k$ to $X_i$ (including the influence of $X_i$ on itself), and the last term is the effect of noise (taking stochastic effects into account).

\section{The Charney-Straus (CS) model}
\label{CSmodel}

The original equations to describe the dynamics of a 2-layer atmosphere proposed by \cite{Charney1980} on a rectangular domain in the extra-tropics are
\begin{eqnarray}
& &\frac{\partial \nabla^2 \psi_1}{\partial t} + (\vec{v}_{1} \cdot \vec{\nabla}) (\nabla^2 \psi_1) + \beta \frac{\partial \psi_1}{\partial x}  =  f_0   \frac{\omega_2}{\delta p} -\frac{K}{\Delta p^2} (\nabla^2 \psi_1 - \nabla^2 \psi_3) \nonumber \\
& & \frac{\partial \nabla^2 \psi_3}{\partial t} + (\vec{v}_{3} \cdot \vec{\nabla}) (\nabla^2 \psi_3 + \frac{f_0 \rho g H}{\Delta p}) + \beta \frac{\partial \psi_3}{\partial x}  =  - f_0   \frac{\omega_2}{\Delta p} + \frac{K}{\Delta p^2} (\nabla^2 \psi_1 - \nabla^2 \psi_3) - C_D \nabla^2 \psi_3 \nonumber \\
& & \frac{\partial }{\partial t} \frac{\psi_1 -\psi_3}{\Delta p}+ \vec{v}_{2} \cdot \vec{\nabla} \frac{\psi_1 -\psi_3}{\Delta p} - \frac{\omega_2 \sigma}{f_0}  =  Q
\label{CS1}
\end{eqnarray}

where $\psi_{1,3}$ are the streamfunction at levels 250~hPa and 750~hPa, respectively, $\vec{v}_{1,2,3}$, the horizontal velocities at levels 250~hPa, 500~hPa and 750~hPa, respectively; $\omega_2$ is the generalized vertical velocity in pressure coordinates at 500~hPa; $\frac{\psi_1 -\psi_3}{\Delta p}$ is the vertical discretization of the hydrostatic balance in pressure coordinates; $\frac{\partial \psi}{\partial p} = \frac{1}{\rho f_0} = \frac{RT}{p f_0}$ where $\rho$ is the density, $T$ the temperature and $R$ the perfect gaz constant; $\sigma$ is the static stability; $H$ is the height of the orography; $f_0$ is the Coriolis parameter at $\phi_0=45^\circ$; $\Delta p$ is the pressure difference between the two layers; $C_D$ is the surface friction coefficient; $K$ is the friction between the two atmospheric layers related to the vertical Reynolds Stress; and $Q=\gamma ((\psi_1-\psi_3)^* - (\psi_1-\psi_3))$ is the Newtonian cooling where $(\psi_1-\psi_3)^*$ is a reference climatology around which the solution of the system oscillates. 

These equations are made non-dimensional by rescaling time, space and the fields as follows: $t' = tf_0$, $ x'=\frac{x}{L}$, $y'=\frac{y}{L}$, $\psi'=\frac{\psi}{L^2 f_0}$, $\omega'=\frac{\omega}{f_0 \Delta p}$. The key parameters are also made non-dimensional as: $H'=\frac{H \rho g}{\Delta p}$, $\beta^*=\frac{\beta L}{f_0}$, $k'=\frac{K}{f_0 \Delta^2 p}$, $\sigma'=\frac{\sigma \Delta^2 p}{L^2 f_0^2} $, $h"=\frac{\gamma \Delta p}{f_0}$, and $C'=\frac{C}{f_0}$. The equations can then be written as

\begin{eqnarray}
\frac{\partial \nabla^2 \psi'_1}{\partial t'} & + &(\vec{v'}_{1} \cdot \vec{\nabla}) (\nabla^2 \psi'_1) + \beta^* \frac{\partial \psi'_1}{\partial x'}  =   \omega'_2 - k' (\nabla^2 \psi'_1 - \nabla^2 \psi'_3) \nonumber \\
\frac{\partial \nabla^2 \psi'_3}{\partial t'}  & + &(\vec{v'}_{3} \cdot \vec{\nabla}) (\nabla^2 \psi'_3 + H') + \beta^* \frac{\partial \psi'_3}{\partial x'}  = - \omega_2' + k' (\nabla^2 \psi'_1 - \nabla^2 \psi'_3) - C' \nabla^2 \psi'_3 \nonumber \\
\frac{\partial }{\partial t'} (\psi'_1 -\psi'_3) & + & \vec{v'}_{2} \cdot \vec{\nabla} (\psi'_1 -\psi'_3) - \omega'_2 \sigma' =  h" ((\psi'_1-\psi'_3)^* - (\psi'_1-\psi'_3))
\label{CS1-adim}
\end{eqnarray}

These equations can be further simplified by summing and differentiating the two first equations, and replacing  $\omega'_2$ by the third one. This leads to two equations for the evolution of the barotropic $\psi=\frac{\psi'_1+\psi'_3}{2}$ and the baroclinic $\theta=\frac{\psi'_1-\psi'_3}{2}$ streamfunctions. 

The domain over which the equations are integrated is rectangular with an aspect ratio $n=\frac{2 L_y}{L_x}$ where $L_x$ and $L_y$ are the zonal and meridional lengths of the domain. $L_y$ is further related to the characteristic space scale $L$ used for scaling as $L_y= \pi L$.  

The non-dimensional fields can then be developed in Fourier series, and severely truncated to the first dominant modes. \cite{Charney1980} choose as basis functions for building their low-order model: 
\begin{align*}
&F_A = \sqrt{2} \cos(y') \\
&F_K = 2 \cos(n x') \sin(y') \\
&F_L = 2  \sin(n x') \sin(y') \\
&F_C = \sqrt{2} \cos(2y') \\
&F_M = 2  \cos(n x') \sin(2y') \\
&F_N = 2 \sin(n x') \sin(2y')
\end{align*}
and each field is then written as: $ \psi = \psi_A F_A + \psi_K F_K + \psi_L F_L + \psi_C F_C + \psi_M F_M + \psi_N F_N$ and $ \theta = \theta_A F_A + \theta_K F_K + \theta_L F_L + \theta_C F_C + \theta_M F_M + \theta_N F_N$. The coefficients are normalized, leading to orthonormal modes in a $L_2$ norm. They further assume that the thermodynamic climatology $\theta* = \theta_A^* F_A$, and that the orography is defined as $H'=h_k F_K$. 

These fields are then injected into Eq.~(\ref{CS1-adim}) and the equations are projected using the $L_2$ norm on the different modes. This leads to the set of 12 ordinary differential equations given in Appendix A. The parameter values used in the following for this set of equations are provided in Appendix A. 

\subsection{Basic characteristics of the solutions}

The solutions of this model have been extensively explored in different contexts. First, \cite{Charney1980} analyzed the emergence of zonal and blocking flows and the role of the orography in the emergence of instabilities, competing with the classical baroclinic instablity. Note that subsequent works, by extending the number of modes, have criticized the emergence of zonal and blocking flows as well as the mechanisms associated with this emergence, but this question is beyond the scope of the current analysis. For more information on this topic, the reader is directed toward \cite{Cehelsky1987} and the recent paper of \cite{Xavier2024}. The model is however a good candidate for different types of analyses as it shows similarities with the type of dynamics that can be found in the atmosphere and is easy to implement, run and analyze. An example is provided in \cite{Demaeyer2020}, in which a methodology is proposed to modify statistical post-processing schemes when model parameters change.

\begin{figure*}
    \centering
    \includegraphics{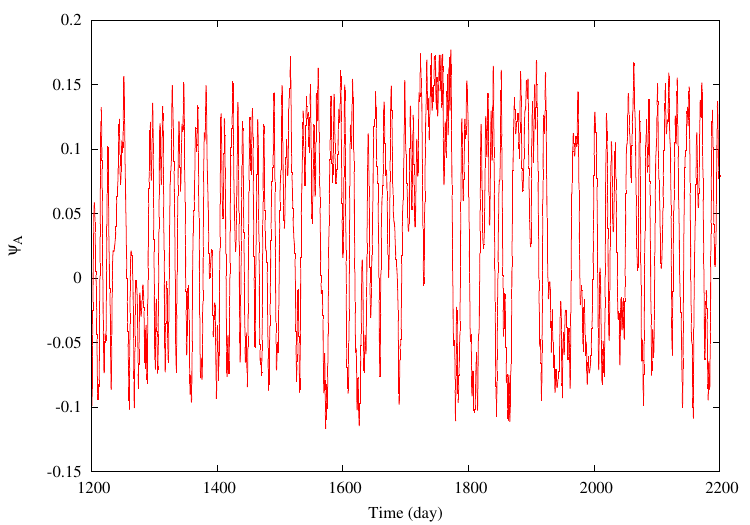}
    \caption{Temporal evolution of the variable $\psi_A$.}
    \label{fig:solution}
\end{figure*}

With the parameter values given in Appendix A, a solution is generated for the model using a second-order Heun integration scheme with a time step of $\Delta t=0.01$ time unit. Figure~\ref{fig:solution} displays the temporal evolution of $\psi_A$ generated by this deterministic dynamical system. It clearly displays an erratic behavior reminiscent of the type of behavior one experiences in the real atmosphere. This erratic behavior is also reminiscent of the chaotic nature of the dynamics, which means that the solution displays the property of sensitivity to initial conditions. 

\begin{figure*}
    \centering
    \includegraphics{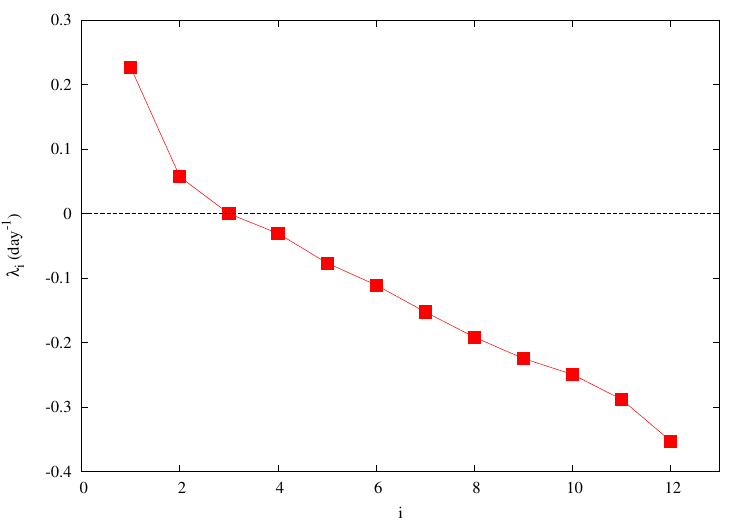}
    \caption{Lyapunov spectrum of the solution generated by the CS model described in Appendix A.}
    \label{fig:lyapunov}
\end{figure*}

This property of sensitivity to initial conditions may be checked by computing the Lyapunov exponents, as for instance discussed in \cite{Eckmann1985}, \cite{Parker1989}, \cite{Nicolis1995}, and \cite{Vannitsem2017}. The Lyapunov exponents characterize the exponential sensitivity to initial condition errors and their number is equal to the dimension of the system. If positive Lyapunov exponents are present, the system displays sensitivity to initial conditions, and hence a chaotic dynamics. Figure~\ref{fig:lyapunov} displays the Lyapunov spectrum, i.e. the ensemble of exponents ranked from the largest to the smallest, of the solution generated by the Charney-Straus (CS) model. There are two positive exponents, indicating that the solution of the CS model for the chosen parameters is chaotic. The dominant exponent has a value of $0.23$ day$^{-1}$, leading to a doubling time of errors of the order of $3$ days, a typical value for more sophisticated systems \citep{Vannitsem2017}.

This chaotic behavior is reminiscent of the sensitivity to initial conditions found in the real atmosphere, as illustrated in \cite{Lorenz1969} or in more realistic atmospheric models like the one studied in \cite{Lorenz1982}, \cite{Dalcher1987}, and \cite{Savijarvi1995}. This makes the current solution a good candidate for an analysis of the uncertainty propagation through the dynamics of the Shannon entropy. Note that for the computation of the different terms of the entropy budget in the following section, 5,000,000 time units are used (approximatively 550,000 days). 

\section{The entropy budget}
\label{Results}

In this section, the computation of the different terms discussed in Section \ref{2.1} on the dynamics of the CS model presented in Section \ref{CSmodel}, is discussed. We start with the single RETs as these are the usual terms that are computed for characterizing the dependencies from one variable to another. We then present the entropy balance terms, i.e. the SEG and the total RETs, together with the different (additive) contributions associated with the physical terms. The section is then closed with a synthesis of the different contributions to the total RET, focusing on the nonlinear terms. Conditional expectations are obtained from a coarse-grained binning of 50 equal-amplitude bins, and derivatives are estimated by central differences.

\subsection{Single RETs}

Let us start by investigating what can be learned from the single RETs. Here, there is no decomposition in terms of the different physical processes as, usually, the goal is to investigate the overall influence of a variable on another. In Table~\ref{tab:transfer-lin}, each column (from 2 to 12) represents the influence of a variable on the target variable listed in the first column of the table. 

One may split the variables into two blocks, the first group, composed of variables $\psi_A, \theta_A, \psi_K, \theta_K, \psi_L, \theta_L$, which are the first meridional modes, and the second group, including variables $\psi_C, \theta_C, \psi_M, \theta_M, \psi_N, \theta_N$, which are the second meridional modes. The influence of variables in the first group on variables in this same group dominates as compared with the influence of variables in the second group on variables in the first group. This reflects the weaker influence of the second group on the first one. However, as we will see in Section~\ref{global_entropy}, synergetic terms of the second group are relatively high, even if less dominant compared to the synergetic terms of the first group of variables (Table~\ref{tab:Synergy}). So, the single RETs do not provide the most accurate picture of the influence of the second group on the first one. 

If we focus on the second group as a target, the dominant influence still occurs within the group, but some key RETs from the first group are important. For instance the influence of $\psi_L$ on $\theta_M$ if of the same order as the influence among the variables within the second group, suggesting the relatively important influence of the first group on the second one. This however should again be tempered by the fact that synergetic terms that are essential components of the change of information entropy, are not accounted for here either. 

\begin{sidewaystable*}
	\centering
 \setlength{\tabcolsep}{1pt}
	\caption{Single RETs between the different variables of the CS model. The first column indicates the target variable, while the subsequent columns contain the single RETs associated with the influence of variables listed in the first row.}
	\begin{tabular}{l|rr|rr|rr|rr|rr|rr|}
		\hline 
		Target & $\psi_A$ & $\theta_A$ & $\psi_K$ & $\theta_K$ & $\psi_L$ & $\theta_L$ & $\psi_C$ & $\theta_C$ & $\psi_M$ & $\theta_M$ & $\psi_N$ & $\theta_N$ \\
		\hline
	      $\psi_A$ &  - &  $3.8 \, 10^{-4}$ & 0 & 0 & $1.12 \, 10^{-2}$ & $-5.93 \, 10^{-3}$ & 0 & 0 & 0 & 0 & 0 & 0 \\ 
               $\theta_A$ & $1.57 \, 10^{-3}$ & - & $4.84 \, 10^{-4}$ & $-3.02 \, 10^{-4}$& $3.15 \, 10^{-4}$ &$-1.61 \, 10^{-3}$ & 0 & 0 & $2.9 \, 10^{-7}$ &  $-3.49 \, 10^{-7}$ &  $-3.9 \, 10^{-7}$ & $2.03 \, 10^{-7}$ \\
        $\psi_K$ & $8.9 \, 10^{-4}$  & $-1.98 \, 10^{-3}$ & -  & $1.58 \, 10^{-3}$ & $ -4.62 \, 10^{-5} $ & $1.04 \, 10^{-2} $ & $2.95 \, 10^{-7} $  &  $ -4.92 \, 10^{-8} $ & 0 & 0 & $ -2.63 \, 10^{-7} $  & $ -1.51 \, 10^{-7} $ \\

        $\theta_K$ & $4.23 \, 10^{-3}$ & $-4.84 \, 10^{-3}$ & $5.463 \, 10^{-3}$& - & $2.69 \, 10^{-2}$ & $-3.08 \, 10^{-3}$ & $-1.41 \, 10^{-6}$ & $6.94 \, 10^{-8}$ & 0 & 0 & $-1.16 \, 10^{-7}$ & $1.71 \, 10^{-7}$ \\
        $\psi_L$ &  $-5.51 \, 10^{-4}$ & $-2.25 \, 10^{-3}$ & $4.87 \, 10^{-4}$ & $9.65 \, 10^{-3}$ & - & $1.48 \, 10^{-3}$ & $6.53 \, 10^{-7}$ & $3.61 \, 10^{-8}$& $6.94 \, 10^{-7}$ & $5.05 \, 10^{-8}$ & 0  & 0 \\
        $\theta_L$ &  $4.49 \, 10^{-4}$ & $-5.04 \, 10^{-3}$ & $2.044 \, 10^{-2}$ & $2.25 \, 10^{-3}$ & $6.26 \, 10^{-3}$ & - & $4.70 \, 10^{-7}$ & $-9.54 \, 10^{-9}$ & $2.06 \, 10^{-7}$ & $-7.93 \, 10^{-7} $ & 0 & 0 \\
        $\psi_C$ & 0 & 0 & $4.06 \, 10^{-4}$ & $-1.80 \, 10^{-4}$ & $-4.61 \, 10^{-4}$ & $-9.04 \, 10^{-5}$ & - & $1.45 \, 10^{-3}$ &   $1.15 \, 10^{-3}$ & $2.09 \, 10^{-4}$ & $-1.87 \, 10^{-3}$ &  $-2.45 \, 10^{-4}$ \\
        $\theta_C$ & 0 & 0 & $1.82 \, 10^{-3}$ & $8.26 \, 10^{-5}$ & $4.46 \, 10^{-4}$ & $6.45 \, 10^{-5}$ & $7.46 \, 10^{-3}$ & - & $-1.13 \, 10^{-4}$ & $-1.21 \, 10^{-4}$ & $3.88 \, 10^{-3}$ &  $-2.03 \, 10^{-3}$ \\
        $\psi_M$ & $-1.89 \, 10^{-4}$ & $5.99 \, 10^{-5}$ & 0 & 0 & $2.71 \, 10^{-5}$ & $4.68 \, 10^{-6}$ & $1.08 \, 10^{-5}$ & $1.16 \, 10^{-6}$ & - & $1.85 \, 10^{-3}$ & $5.12 \, 10^{-4}$ & $6.3 \, 10^{-5}$ \\
        $\theta_M$ & $7.70 \, 10^{-5}$& $1.94 \, 10^{-5}$ & 0 & 0 & $-1.47 \, 10^{-3}$ & $5.43 \, 10^{-4}$ & $4.3 \, 10^{-3}$& $-1.01 \, 10^{-3}$ & $5.07 \, 10^{-3}$ & -  &  $-3.41 \, 10^{-5}$ & $-9.90 \, 10^{-5}$ \\
        $\psi_N$ & $2.42 \, 10^{-4}$ & $-4.26 \, 10^{-6}$ & $-9.48 \, 10^{-5}$ & $-9.26 \, 10^{-6}$ & 0 & 0 & $1.25 \, 10^{-2}$ & $-4.08 \, 10^{-3}$ & $-6.57 \, 10^{-4}$ &  $-1.29 \, 10^{-5}$ & - & $1.84 \, 10^{-3}$ \\
        $\theta_N$ & $8.18 \, 10^{-5}$ & $1.35 \, 10^{-5}$ & $-1.70 \, 10^{-4}$ & $6.47 \, 10^{-4}$ & 0 & 0 & $3.39 \, 10^{-3}$ & $-4.03 \, 10^{-3}$ & $-4.53 \, 10^{-4}$ & $1.75 \, 10^{-4}$ & $4.69 \, 10^{-3}$ & -  \\
		\hline
	\end{tabular} \label{tab:transfer-lin}

\end{sidewaystable*}

\subsection{Global entropy budget}
\label{global_entropy}

The global entropy budget, which should be equal to 0, is displayed in Table~\ref{tab:entropy-global} for each target variable in the first column. The second column displays the part of the budget associated with the SEG (self-entropy generation; see Eq.~(\ref{SEG})). These quantities are easily computed as the terms proportional to the variables themselves are linear with a friction coefficient proportional to $-k$, associated with the friction term in Eqs. (\ref{a1}-\ref{a12}). The SEG terms are all negative and reflect the dissipative character of the dynamics. The third column displays the total RET from all variables to the target variable (Eq.~(\ref{decompositionentropy})). The fourth column represents the contributions of the sum of the single RETs, and the fifth column displays the contribution of the synergetic terms, $T_{Syn}$ in Eq.~(\ref{Tsyn}), reflecting the co-variability of the variables appearing in the right-hand side of all the equations given in Appendix A (\ref{a1}-\ref{a12}). Note that the balance between the SEG (second column) and the total RET (third column) is close to be met, with a slight imbalance that could possibly be due to different reasons: the numerical round-off errors, the sample used, and/or the nature of the dynamics of the system which is deterministic and chaotic without stochastic components. The latter implies that the probability density function of this system could be non-differentiable in certain directions, an important condition for computing the information entropy budget. This interesting question is out of scope of the present work and will be addressed in the future.

\begin{table*}
	\centering
 \setlength{\tabcolsep}{3pt}
	\caption{Global entropy budget: The first column contains the name of the target variable, the second column the self-entropy generation term (SEG), the third column the total rate of entropy transfer (RET) from all variables except the target, the fourth column the sum of all single RETs, and the last column the synergy term, i.e. the difference between the total RET minus the sum of single RETs.}
	\begin{tabular}{l|r|r|r|r|}
		\hline 
	   Target & SEG & Total RET  & Sum of single RETs & Synergy \\
        variable & $E(\partial F_i / \partial X_i)$ &  $ T(X_{ \neq i} \rightarrow X_i)$ & $\sum_{{j \neq i}} T(X_{j} \rightarrow X_i)$ & $T_{Syn}(X_{ \neq i} \rightarrow X_i)$ \\
		\hline
	      $\psi_A$ & $-5.70 \, 10^{-3}$ & $5.68 \, 10^{-3}$ & $5.68 \, 10^{-3}$ & 0. \\ 
               $\theta_A$ & $-1.43 \, 10^{-2}$  &  $1.42 \, 10^{-2}$ & $4.55 \, 10^{-4}$ & $1.38 \, 10^{-2}$ \\
        $\psi_K$ & $-5.70 \, 10^{-3}$  & $5.69 \, 10^{-3}$ & $1.09 \, 10^{-2}$ & $-5.23 \, 10^{-3}$\\

        $\theta_K$ & $-1.91 \, 10^{-2}$  & $1.91 \, 10^{-2}$ & $2.87 \, 10^{-2}$ & $-9.57 \, 10^{-3}$ \\
        $\psi_L$ & $-5.70 \, 10^{-3}$   & $5.69 \, 10^{-3}$ &  $8.82 \, 10^{-3}$ & $-3.12 \, 10^{-3}$ \\
        $\theta_L$ & $-1.91 \, 10^{-2}$  & $1.91 \, 10^{-2}$ &  $2.44 \, 10^{-2}$ & $-5.27 \, 10^{-3}$ \\
        $\psi_C$ & $-5.70 \, 10^{-3}$  & $5.68 \, 10^{-3}$ &$3.69 \, 10^{-4}$ & $5.31 \, 10^{-3}$\\
        $\theta_C$ &  $-1.90 \, 10^{-2}$ & $1.90 \, 10^{-2}$ & $1.15 \, 10^{-2}$  & $7.48 \, 10^{-3}$ \\
        $\psi_M$ & $-5.70 \, 10^{-3}$  & $5.68 \, 10^{-3}$ &  $ 2.34 \, 10^{-3}$ & $3.34 \, 10^{-3}$\\
        $\theta_M$ & $-2.15 \, 10^{-2}$ & $2.14 \, 10^{-2}$ & $7.40 \, 10^{-3}$ & $1.40 \, 10^{-2}$\\
        $\psi_N$ & $-5.70 \, 10^{-3}$  & $5.68 \, 10^{-3}$ & $9.76 \, 10^{-3}$ & $-4.08 \, 10^{-3}$\\
        $\theta_N$ & $-2.15 \, 10^{-2}$ &  $2.14 \, 10^{-2}$ & $4.35 \, 10^{-3}$ & $1.71 \, 10^{-2}$\\
		\hline
	\end{tabular} \label{tab:entropy-global}

\end{table*}

Table~\ref{tab:entropy-global} shows the absence of synergetic terms influencing variable $\psi_A$ as there are only linear terms influencing this variable, see Eq.~(\ref{a1}). On the contrary, synergetic terms dominate $\theta_A$, indicating that the nonlinear co-variability is key in the change of information entropy for this variable (see Eq.~(\ref{a4})). Synergetic terms also dominate for variable $\psi_C$, but to a lesser extent. For the other variables, both synergetic and single RETs are important in the change of information entropy.

Let us now go more deeply in the specific contributions of the different terms. Table~\ref{tab:entropy-lin} displays the specific contributions of the linear terms to the total RET in the equations related to Earth rotation, orography and friction (excluding the self-influencing term). All the corresponding terms are displayed together with the values obtained along the trajectory of the model solution. Concerning the rotation terms (second and third columns), the dominant contributions are the ones related to the baroclinic modes, $\theta_K$ and $\theta_L$. Note that the sign changes from one variable to another, reflecting a tendency to control or to amplify the variability of the target variable (displayed in the first column).

\begin{table*}
	\centering
 \setlength{\tabcolsep}{3pt}
	\caption{Contributions of the linear terms to the entropy budget. The first column represents the target variable. The following columns contain the linear terms of the evolution equations and their respective value (see Appendix A): the Coriolis term on a beta-plane referred as rotation, the orographic term, and the friction term.}
	\begin{tabular}{l||r|r||r|r||r|r||}
		\hline 
		Target & Rotation &  Value & Orography & Value & Friction & Value  \\
		\hline
	      $\psi_A$ & - & - & $  0.5 h_{01} (\psi_L-\theta_L) $ & $5.27 \, 10^{-3}$ & $ k \theta_A$ &  $3.80 \, 10^{-4}$ \\ 
        \hline
        $\theta_A$ &  - & - & $-0.5 h_{01} (\psi_L-\theta_L) / (1+\frac{1}{\sigma_0})$ & $-1.51  \, 10^{-3}$ & $k \psi_A  / (1+\frac{1}{\sigma_0}) $&  $1.57 \, 10^{-3}$ \\
        \hline
        $\psi_K$ & $\beta_1 \psi_L $ & $-2.04 10^{-4} $ & - & - & $k \theta_K $ &  $1.58 \, 10^{-3}$ \\
        \hline
        $\theta_K$ &  $\beta_1 \theta_L / (1+\frac{1-\beta}{\sigma_0})$ & $1.18 \, 10^{-3}$ & -  & - & $ k \psi_K /(1+\frac{1-\beta}{\sigma_0})$ &  $5.46 \, 10^{-3}$ \\
        \hline
        $\psi_L$ &  $-\beta_1 \psi_K$ &  $7.92 \, 10^{-4}$ & $-0.5 h_{n1} (\psi_A - \theta_A) $ & $-8.19 10^{-4}$ & $ k \theta_L$ & $1.48 \, 10^{-3}$ \\
        \hline
        $\theta_L$ & $-\beta_1 \theta_K /(1+\frac{1-\beta}{\sigma_0})$ & $-1.47 \, 10^{-3}$ & $0.5 h_{n1} (\psi_A - \theta_A) / (1+\frac{1-\beta}{\sigma_0})$ & $3.70 \, 10^{-3}$ & $k \psi_L  (1+\frac{1-\beta}{\sigma_0}) $& $6.26 \, 10^{-3}$ \\
        \hline
        $\psi_C$ & -  & - & $0.5 h_{02} (\psi_N - \theta_N) $ & $-1.39 \, 10^{-3}$ & $ k \theta_C$ &    $1.45 \, 10^{-3}$   \\
        \hline
        $\theta_C$ & -  & - & $-0.5 h_{02} (\psi_N - \theta_N) / (1+\frac{1-\epsilon}{\sigma_0})$ & $4.81 10^{-3}$ & $ k \psi_C/ (1+\frac{1-\epsilon}{\sigma_0})$ & $7.46 \, 10^{-3}$\\
        \hline
        $\psi_M$ & $\beta_2 \psi_N $  & $-1.92 \, 10^{-4}$ & -  &-  & $ k \theta_M $ &  $1.85 \, 10^{-3}$\\
        \hline
        $\theta_M$ & $\beta_2 \theta_N / (1+\frac{1-\beta'}{\sigma_0})$ & $6.62 \, 10^{-7}$ & - & - & $ k \psi_M/ (1+\frac{1-\beta'}{\sigma_0})$ & $5.07 \, 10^{-3}$\\
        \hline
        $\psi_N$ &  $-\beta_2 \psi_M $ & $2.33 \, 10^{-4}$ & $-0.5 h_{n2} (\psi_C - \theta_C) $ & $8.75 \, 10^{-3}$& $ k \theta_N $ & $1.84 \, 10^{-3}$\\
        \hline
        $\theta_N$ & $-\beta_2 \theta_M / (1+\frac{1-\beta'}{\sigma_0})$ & $-1.23 \, 10^{-4}$  & $0.5 h_{n2} (\psi_C - \theta_C) / (1+\frac{1-\beta'}{\sigma_0})$ & $5.07 \, 10^{-3}$& $ k \psi_N/ (1+\frac{1-\beta'}{\sigma_0})$ & $4.69 \, 10^{-3}$\\
		\hline
	\end{tabular} \label{tab:entropy-lin}

\end{table*}

In the fourth and fifth columns, the contributions from the orography are displayed, with an amplitude of the contributions similar to the ones associated with rotation. These quantities are obviously dependent on the imposed height of the mountains. In the absence of mountains, the contributions to the total RETs are equal to 0. Note that the orography can also have a controling effect (negative signs) on the variability of certain variables, either baroclinic or barotropic. Finally, in the sixth and seventh columns, the quantities related to friction that are not associated with the self-influence are displayed. They are all positive, being about $10-50\%$ of the self-influence in absolute value, indicating that their effect is to increase the information entropy, and hence the uncertainty on the variables. 

Let us now turn to the contribution of the nonlinear terms to the total RET. These are displayed in Tables \ref{tab:entropy-nl} and \ref{tab:Synergy}. The former displays the total contribution of the nonlinear terms to the entropy budget, while the latter displays the decomposition of the single RET and synergetic contributions. 

\begin{table*}
	\centering
 \setlength{\tabcolsep}{3pt}
	\caption{Contributions of the nonlinear terms to the total RET. NL1 and NL2 refer to the two sets of nonlinearities, split in terms of variables interacting with each other. For the baroclinic modes, these are also decomposed in a term coming from the vorticity equations and one from the thermodynamic equations, displayed in the first and second rows in the boxes, respectively.}
	\begin{tabular}{l||r|r||r|r||}
		\hline 
		Target &  NL1 & Value & NL2 & Value \\
		\hline
	      $\psi_A$  & - &- &- &- \\ 
        \hline
        $\theta_A$ & $-\frac{\alpha}{\sigma_0} (\theta_K \psi_L - \theta_L \psi_K) / (1+\frac{1}{\sigma_0})$ & $1.26 \, 10^{-2}$ & $-\frac{\alpha'}{\sigma_0} (\theta_M \psi_N - \psi_M \theta_N)/ (1+\frac{1}{\sigma_0})$ & $1.57 \, 10^{-3}$\\
        \hline
       $\psi_K$ & $ -\beta \alpha (\psi_L \psi_A + \theta_L \theta_A) $ & $6.01 10^{-3}$ & $ - \delta \alpha" (\psi_N \psi_C + \theta_N \theta_C) $ & $-1.70 10^{-3}$ \\
       \hline
        $\theta_K$  & $ -\beta \alpha (\theta_L \psi_A + \psi_L \theta_A) / (1+\frac{1-\beta}{\sigma_0})$ & $-3.73 \, 10^{-2}$ & $ -\delta \alpha" (\theta_N \psi_C + \psi_N \theta_C) / (1+\frac{1-\beta}{\sigma_0})$ & $-4.02 \, 10^{-3}$\\ 
         & $- \frac{(1-\beta)}{\sigma_0} \alpha (\theta_L \psi_A - \psi_L \theta_A) / (1+\frac{1-\beta}{\sigma_0})$ & $5.93 \, 10^{-2}$ & $- \frac{(1-\beta)}{\sigma_0} \alpha" (\theta_N \psi_C - \psi_N \theta_C) / (1+\frac{1-\beta}{\sigma_0})$ & $-5.48 \, 10^{-3}$\\
        \hline
        $\psi_L$ &  $\beta \alpha (\psi_K \psi_A + \theta_K \theta_A) $ & $9.09 10^{-3}$ & $\delta \alpha" (\psi_M \psi_C + \theta_M \theta_C)$ & $-4.86 \, 10^{-3}$ \\
        \hline
        $\theta_L$  & $\beta \alpha (\theta_K \psi_A + \psi_K \theta_A) / (1+\frac{1-\beta}{\sigma_0})$  & $-2.97 \, 10^{-2}$  & $\delta \alpha" (\theta_M \psi_C + \psi_M \theta_C) / (1+\frac{1-\beta}{\sigma_0})$  & $-5.41 \, 10^{-3}$\\
          & $ +\frac{(1-\beta)}{\sigma_0} \alpha (\theta_K \psi_A - \psi_K \theta_A) / (1+\frac{1-\beta}{\sigma_0})$ &  $4.98 \, 10^{-2}$& $ +\frac{(1-\beta)}{\sigma_0} \alpha" (\theta_M \psi_C - \psi_M \theta_C) / (1+\frac{1-\beta}{\sigma_0})$ & $-4.09 10^{-3}$ \\
        \hline
        $\psi_C$ & $\epsilon \alpha" (\psi_N \psi_K + \theta_N \theta_K) $ & $1.39 \, 10^{-4}$ & $-\epsilon \alpha" (\psi_M \psi_L + \theta_M \theta_L) $ &    $5.48  \, 10^{-3}$  \\
        \hline
        $\theta_C$ &  $\epsilon \alpha" (\psi_N \theta_K + \theta_N \psi_K)/ (1+\frac{1-\epsilon}{\sigma_0})$ &  $2.75 \, 10^{-3}$ & $-\epsilon \alpha" (\psi_M \theta_L + \psi_M \theta_L) / (1+\frac{1-\epsilon}{\sigma_0}) $ &  $3.98 \, 10^{-3}$ \\
        & $ -\frac{(1-\epsilon)}{\sigma_0} \alpha" (\theta_K \psi_N - \psi_K \theta_N) / (1+\frac{1-\epsilon}{\sigma_0})$ & $-1.51 \, 10^{-3}$& $ -\frac{(1-\epsilon)}{\sigma_0} \alpha" (\theta_M \psi_L - \psi_M \theta_L) / (1+\frac{1-\epsilon}{\sigma_0})$ & $1.47 \, 10^{-3}$ \\
        \hline
        $\psi_M$ & $-\beta' \alpha' (\psi_N \psi_A + \theta_N \theta_A) $  & $2.93 \, 10^{-3}$ & $-\delta' \alpha" (\psi_L \psi_C + \theta_L \theta_C)$  & $1.09 \, 10^{-3}$  \\
        \hline
        $\theta_M$ & $-\beta' \alpha' (\theta_N \psi_A + \psi_N \theta_A) / (1+\frac{1-\beta'}{\sigma_0})$  & $1.43 \, 10^{-3}$& $-\delta' \alpha" (\theta_L \psi_C + \psi_L \theta_C)/ (1+\frac{1-\beta'}{\sigma_0})$ & $9.33 \, 10^{-4}$  \\
         & $ -\frac{(1-\beta')}{\sigma_0} \alpha' (\theta_N \psi_A - \psi_N \theta_A) / (1+\frac{1-\beta'}{\sigma_0})$ & $-3.24 \, 10^{-3}$ & $ -\frac{(1-\beta')}{\sigma_0} \alpha" (\theta_L \psi_C - \psi_L \theta_C) / (1+\frac{1-\beta'}{\sigma_0})$ & $1.72 \, 10^{-2}$ \\
        \hline
        $\psi_N$ &  $\beta' \alpha' (\psi_M \psi_A + \theta_M \theta_A) $  &  $-4.71 \, 10^{-3}$ & $\delta' \alpha" (\psi_K \psi_C + \theta_K \theta_C)$ & $-4.26 \, 10^{-4}$ \\
        \hline
        $\theta_N$ & $\beta' \alpha' (\theta_M \psi_A + \psi_M \theta_A) / (1+\frac{1-\beta'}{\sigma_0})$  & $6.27 \, 10^{-4}$& $\delta' \alpha" (\theta_K \psi_C + \psi_K \theta_C)/ (1+\frac{1-\beta'}{\sigma_0})$ & $4.14 10^{-4}$  \\
         & $ +\frac{(1-\beta')}{\sigma_0} \alpha' (\theta_M \psi_A - \psi_M \theta_A) / (1+\frac{1-\beta'}{\sigma_0})$ & $1.63 \, 10^{-3}$& $ +\frac{(1-\beta')}{\sigma_0} \alpha" (\theta_K \psi_C - \psi_K \theta_C) / (1+\frac{1-\beta'}{\sigma_0})$ &  $9.10 \, 10^{-3}$ \\
		\hline
	\end{tabular} \label{tab:entropy-nl}

\end{table*}

In Table \ref{tab:entropy-nl}, the nonlinear terms NL1 and NL2 refer to the two sets of nonlinearities, separated by the specific variables interacting with each other. Those corresponding to the baroclinic modes are further decomposed in a term coming from the vorticity equations and one from the thermodynamic equations through the vertical velocity $\omega_2$. For the first group of variables ($\psi_A, \theta_A, \psi_K, \theta_K, \psi_L, \theta_L$), NL1 are nonlinear terms combining variables belonging to the same set of modes, while NL2 corresponding to the target variables of the first group are combinations of the modes of the second group $\psi_C, \theta_C, \psi_M, \theta_M, \psi_N, \theta_N$. Interestingly, the NL1 terms play a prominent role in the change of information entropy content for the first group of variables. The overall contribution of NL1 is to increase the information entropy content for this set of variables. On the other hand, the NL2 contributions are smaller in absolute value compared to the NL1 contributions and mostly negative, with the exception of the contribution to the variable $\theta_A$. So NL2 terms tend to reduce the information entropy content, and hence to control the variability of the corresponding variables.

For the second set of target variables, $\psi_C, \theta_C, \psi_M, \theta_M, \psi_N, \theta_N$, the nonlinear terms NL1 and NL2 are made of interactions between this group of variables and the first set of variables, $\psi_A, \theta_A, \psi_K, \theta_K, \psi_L, \theta_L$. A first remark is that the amplitudes of the NL1 contributions are generally smaller than for the previous group, and that the NL2 terms tend to dominate. In addition, one cannot underline a specific trend in the sign of the contributions within NL1 and NL2 for these targeted variables.

\begin{sidewaystable*}
	\centering
\setlength{\tabcolsep}{3pt}
\caption{Contributions of the nonlinear terms to the entropy budget, decomposed into synergetic and sum of single RET contributions. The contributions to the synergetic terms from each nonlinearity, NL1 and NL2, can be computed thanks to the additive character of Eq.~(\ref{totalret}) and of the single RETs associated with each nonlinear term.}
	\begin{tabular}{l||r|r|r ||r|r|r ||}
		\hline 
		Target &  NL1 & Synergy & Single RET sum & NL2 & Synergy & Single RET sum \\
		\hline
	      $\psi_A$  & - &- &- &- & - & - \\ 
       \hline
      $\theta_A$ & $-\frac{\alpha}{\sigma_0} (\theta_K \psi_L - \theta_L \psi_K) / (1+\frac{1}{\sigma_0})$ & $1.22 \, 10^{-2}$ & $3.90 \, 10^{-4}$ & $-\frac{\alpha'}{\sigma_0} (\theta_M \psi_N - \psi_M \theta_N)/ (1+\frac{1}{\sigma_0})$ & $1.57 \, 10^{-3}$ & $ -6.52 \, 10^{-7} $\\
       \hline
      $\psi_K$ & $ -\beta \alpha (\psi_L \psi_A + \theta_L \theta_A) $ & $-3.54 \, 10^{-3}$ & $9.55 \, 10^{-3}$ & $ - \delta \alpha" (\psi_N \psi_C + \theta_N \theta_C) $ & $-1.70 \, 10^{-3}$ & $ -1.68 \, 10^{-7} $ \\
      \hline
        $\theta_K$  & $ -\beta \alpha (\theta_L \psi_A + \psi_L \theta_A) / (1+\frac{1-\beta}{\sigma_0})$  & $-9.48 \, 10^{-5}$ & $2.21 \, 10^{-2}$ & $ -\delta \alpha" (\theta_N \psi_C + \psi_N \theta_C) / (1+\frac{1-\beta}{\sigma_0})$ & $  -9.50 \, 10^{-3}$ & $ 1.29 \, 10^{-6}$ \\ 
         & $- \frac{(1-\beta)}{\sigma_0} \alpha (\theta_L \psi_A - \psi_L \theta_A) / (1+\frac{1-\beta}{\sigma_0})$ & & & $- \frac{(1-\beta)}{\sigma_0} \alpha" (\theta_N \psi_C - \psi_N \theta_C) / (1+\frac{1-\beta}{\sigma_0})$ & & \\
        \hline
        $\psi_L$ &  $\beta \alpha (\psi_K \psi_A + \theta_K \theta_A) $ &  $1.72 \, 10^{-3}$ & $7.37 \, 10^{-3}$ &  $\delta \alpha" (\psi_M \psi_C + \theta_M \theta_C)$ & $-4.86 \, 10^{-3}$ & $ 1.43 \, 10^{-6}$ \\
        \hline
        $\theta_L$  & $\beta \alpha (\theta_K \psi_A + \psi_K \theta_A) / (1+\frac{1-\beta}{\sigma_0})$  &  $4.24 \, 10^{-3}$ & $1.59 \, 10^{-2}$ & $\delta \alpha" (\theta_M \psi_C + \psi_M \theta_C) / (1+\frac{1-\beta}{\sigma_0})$  & $-9.50 \, 10^{-3}$ & $ 4.45 \, 10^{-7}$ \\
          & $ +\frac{(1-\beta)}{\sigma_0} \alpha (\theta_K \psi_A - \psi_K \theta_A) / (1+\frac{1-\beta}{\sigma_0})$ & & & $ +\frac{(1-\beta)}{\sigma_0} \alpha" (\theta_M \psi_C - \psi_M \theta_C) / (1+\frac{1-\beta}{\sigma_0})$ &  & \\
        \hline
        $\psi_C$ & $\epsilon \alpha" (\psi_N \psi_K + \theta_N \theta_K) $ & $6.47 \, 10^{-4}$& $-5.08 \, 10^{-4}$ & $-\epsilon \alpha" (\psi_M \psi_L + \theta_M \theta_L) $ &   $4.68 \, 10^{-3}$ &  $8.04 \, 10^{-4}$\\
        \hline
        $\theta_C$ &  $\epsilon \alpha" (\psi_N \theta_K + \theta_N \psi_K)/ (1+\frac{1-\epsilon}{\sigma_0})$ & $2.33 \, 10^{-3}$ & $-1.09 \, 10^{-3}$ & $-\epsilon \alpha" (\psi_M \theta_L + \psi_M \theta_L) / (1+\frac{1-\epsilon}{\sigma_0}) $ &  $5.16 \, 10^{-3}$ & $2.83 \, 10^{-4}$\\
        & $ -\frac{(1-\epsilon)}{\sigma_0} \alpha" (\theta_K \psi_N - \psi_K \theta_N) / (1+\frac{1-\epsilon}{\sigma_0})$ & & & $ -\frac{(1-\epsilon)}{\sigma_0} \alpha" (\theta_M \psi_L - \psi_M \theta_L) / (1+\frac{1-\epsilon}{\sigma_0})$ & & \\
        \hline
        $\psi_M$ & $-\beta' \alpha' (\psi_N \psi_A + \theta_N \theta_A) $  & $2.30 \, 10^{-3}$ & $6.37 \, 10^{-4}$ & $-\delta' \alpha" (\psi_L \psi_C + \theta_L \theta_C)$  & $1.05 \, 10^{-3}$ & $4.20 \, 10^{-5}$\\
        \hline
        $\theta_M$ & $-\beta' \alpha' (\theta_N \psi_A + \psi_N \theta_A) / (1+\frac{1-\beta'}{\sigma_0})$  & $-1.79 \, 10^{-3}$ & $-2.60 \, 10^{-5}$ & $-\delta' \alpha" (\theta_L \psi_C + \psi_L \theta_C)/ (1+\frac{1-\beta'}{\sigma_0})$ &  $1.58 \, 10^{-2}$ & $2.34 \, 10^{-3}$\\
         & $ -\frac{(1-\beta')}{\sigma_0} \alpha' (\theta_N \psi_A - \psi_N \theta_A) / (1+\frac{1-\beta'}{\sigma_0})$ & & & $ -\frac{(1-\beta')}{\sigma_0} \alpha" (\theta_L \psi_C - \psi_L \theta_C) / (1+\frac{1-\beta'}{\sigma_0})$ & & \\
        \hline
        $\psi_N$ &  $\beta' \alpha' (\psi_M \psi_A + \theta_M \theta_A) $  & $-4.04 \, 10^{-3}$ & $-6.69 \, 10^{-4}$ & $\delta' \alpha" (\psi_K \psi_C + \theta_K \theta_C)$ & $-2.93 \, 10^{-5}$ & $-3.97 \, 10^{-4}$\\
        \hline
        $\theta_N$ & $\beta' \alpha' (\theta_M \psi_A + \psi_M \theta_A) / (1+\frac{1-\beta'}{\sigma_0})$  &$2.32 \, 10^{-3}$ &  $-6.19 \, 10^{-5}$ & $\delta' \alpha" (\theta_K \psi_C + \psi_K \theta_C)/ (1+\frac{1-\beta'}{\sigma_0})$ & $1.47 \, 10^{-2}$ & $-5.21 \, 10^{-3}$\\
         & $ +\frac{(1-\beta')}{\sigma_0} \alpha' (\theta_M \psi_A - \psi_M \theta_A) / (1+\frac{1-\beta'}{\sigma_0})$ &  & & $ +\frac{(1-\beta')}{\sigma_0} \alpha" (\theta_K \psi_C - \psi_K \theta_C) / (1+\frac{1-\beta'}{\sigma_0})$ &  & \\
		\hline
	\end{tabular} \label{tab:Synergy}

\end{sidewaystable*}

Table \ref{tab:Synergy} shows the decomposition into single RETs and synergetic contribution (related to the combined effect of variables). For the first group of variables ($\psi_A, \theta_A, \psi_K, \theta_K, \psi_L, \theta_L$), the dominant terms are the single RETs of NL1, except for $\theta_A$. For the latter, the dominant term is the synergetic one, indicating the importance of the co-variability of $\psi_L$ and $\theta_K$ in the change of information entropy. For this group of variables, it is also interesting to note that for the thermodynamic nonlinearities NL2, the main contributions are the synergetic ones. For the second group of target variables ($\psi_C, \theta_C, \psi_M, \theta_M, \psi_N, \theta_N$), the trends are less clear but synergetic terms usually dominate. 

These results underline the importance of the co-variability between different variables in driving the dynamics, and in particular, the role of synergies on the change of information entropy. The single RETs only provide a partial diagnostics on the information entropy dynamics, which could provide a biased estimation of causal influences between variables if synergies are not taken into account. This question will be taken up further in the next sections.

\subsection{Summary of the contributions to the total RETs}

\begin{figure*}
    \centering
    \includegraphics[width=10cm]{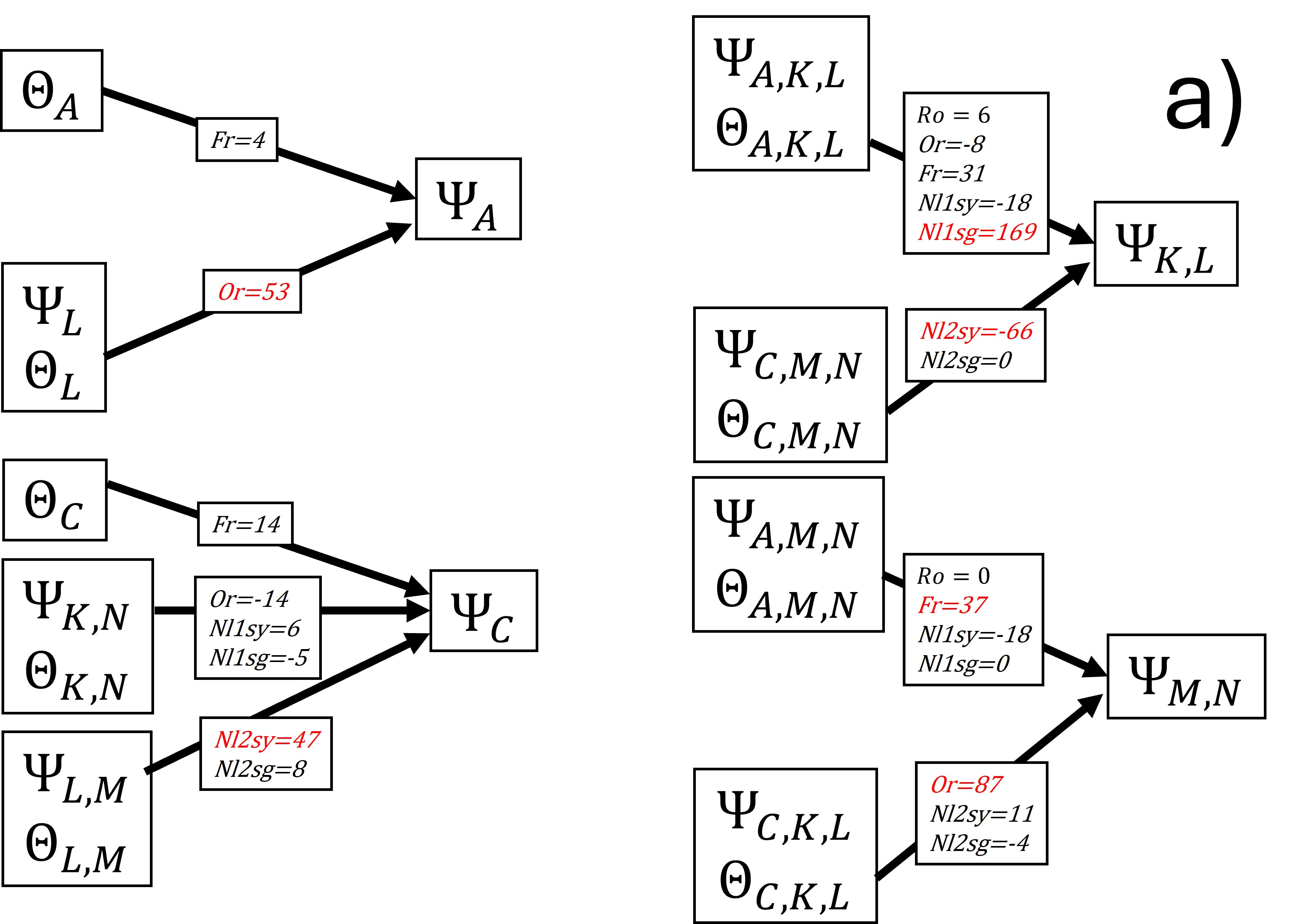}
     \bigskip
      \bigskip
      
    \includegraphics[width=10cm]{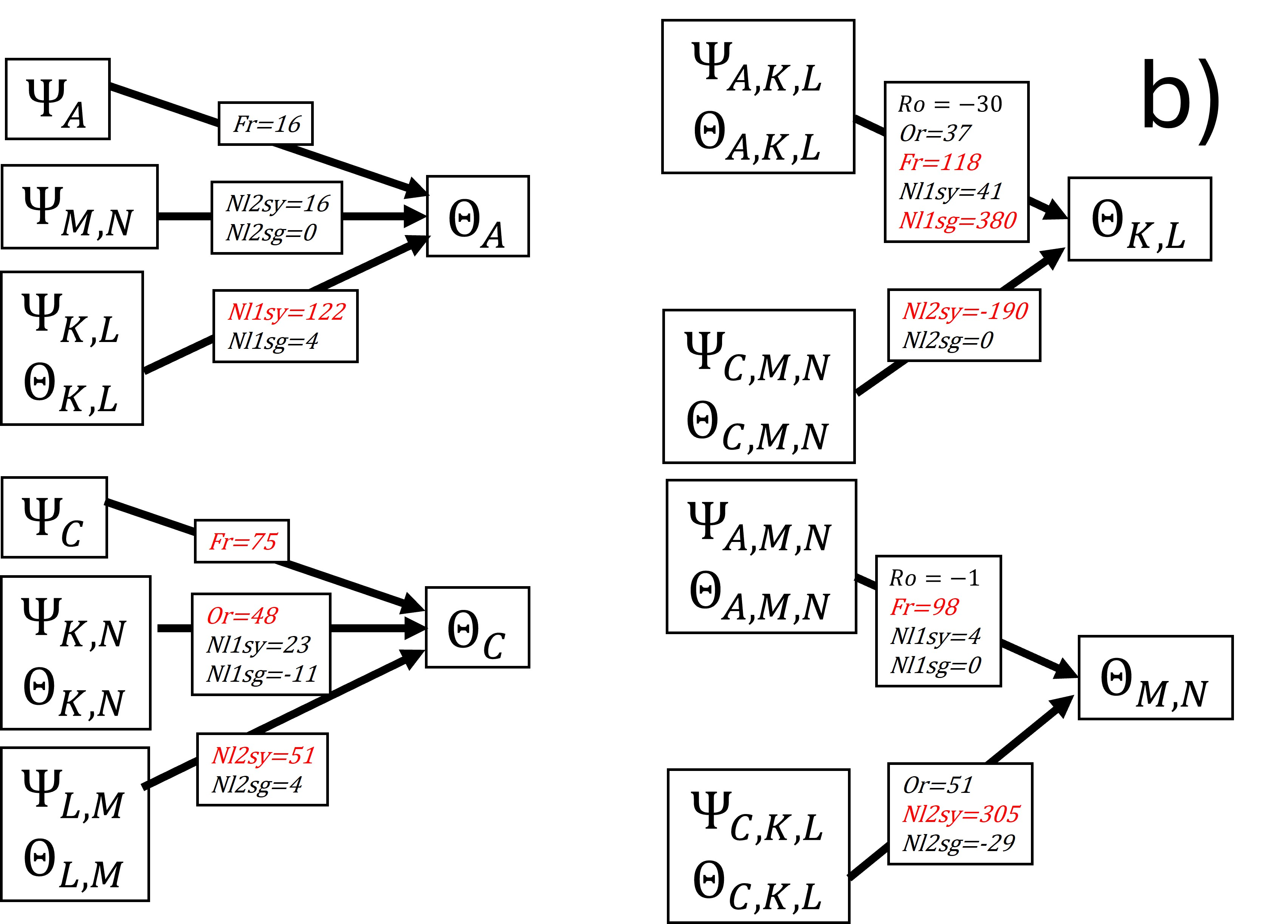}
    \caption{Drivers of the entropy changes for each target variable. Panel (a) displays the barotropic variables: $\psi_A, \psi_{(K,L)}, \psi_C, \psi_{(M,N)}$, while panel (b) shows the baroclinic variables: $\theta_A, \theta_{(K,L)}, \theta_C, \theta_{(M,N)}$. In the eight schemes, the influencing and target groups are on the left and right, respectively. These are connected in the middle by arrows labeled with the budget amplitudes (in  $10^{-4}$ nats/time unit) and the nature of the processes acting on the target variable. The leading terms appear in red.}
    \label{fig:summary}
\end{figure*}

The values and formulas of the entropy budget terms that are evaluated in Tables~\ref{tab:entropy-global}-\ref{tab:Synergy} allow for isolating the main drivers of the rates of entropy change. The amplitudes of these different influences and their origin are synthesized in Fig.~\ref{fig:summary}, in boxes superimposed to the arrows, given in units of $10^{-4}$ nats/time unit. Those amplitudes reflect the importance of specific variables in driving other variables and the leading physical processes at the origin of those influences. The most prominent processes are displayed in red.  In Fig.~\ref{fig:summary}, these processes are referred to as `Ro' (rotation), `Or' (orography), `Fr' (friction), `Nl1sy' and `Nl1sg' (synergetic and single contributions of NL1), and `Nl2sy' and `Nl2sg' (synergetic and single contributions of NL2). The groups of target variables are split into barotropic $\psi$ (Fig.~\ref{fig:summary}a) and baroclinic $\theta$ (Fig.~\ref{fig:summary}b), each one split into zonal/broad (mode A), zonal/sharp (mode C), eddy/broad (modes K,L) and eddy/sharp (modes M,N). The zonal and eddy modes refer to longitude independent and longitude wavy dependent, respectively, whereas the broad and sharp modes refer to latitude dependency with wave number 1 and 2, respectively. The drivers (intervening in the different budget terms) can act as controllers or disturbers, corresponding respectively to negative or positive entropy budget terms. The entropy budget terms concerning a group of target variables are obtained as sums of the respective terms valid for the single target variables in the group. For instance, the NL2sy term of the target group $\psi_{(M,N)}$ and coming from the influencing group $\psi_{(C,K,L)},\theta_{(C,K,L)}$ is the sum of the respective NL2sy terms appearing in the entropy budget of $\psi_M$ and $\psi_N$. 

The analysis of the leading budget terms of barotropic modes (Fig. \ref{fig:summary}a) reveals that the barotropic/zonal/broad mode $\psi_A$ is mostly orographically disturbed by eddy/broad modes, whereas the barotropic/zonal/sharp mode $\psi_C$ is disturbed by the synergy across eddy modes from NL2. The barotropic/eddy/broad modes $\psi_{(K,L)}$ are mostly disturbed by the single contributions of eddy/broad modes in NL1 and controlled by the synergy of eddy/sharp modes of NL2. The barotropic/eddy/sharp modes $\psi_{(M,N)}$ are mostly disturbed by eddy modes through friction and orography.

Similarly for the baroclinic modes (Fig. \ref{fig:summary}b), the analysis reveals that the barotropic/zonal/broad mode $\theta_A$ is mostly disturbed by the synergy of eddy/broad modes through the NL1 terms, whereas the baroclinic/zonal/sharp mode $\theta_C$ is mostly disturbed by the friction from the barotropic/zonal/sharp mode, by the orographic forcing of the eddy modes and by the synergy of eddy modes through NL2 terms.
The baroclinic/eddy/broad modes $\theta_{(K,L)}$ are mostly disturbed by friction coming from eddy/broad modes and controlled by single contributions of  NL2 coming from eddy/sharp modes. Finally, the baroclinic/eddy/sharp modes $\theta_{(M,N)}$  are mostly disturbed by friction coming from eddy/sharp modes and the synergy of eddy/broad modes of NL2.

\section{The Liang's approach based on linear stochastic modelling}
\label{timeseries}

Liang developed an approach to estimate the single RETs from time series \citep{Liang2014,Liang2021}. It relies on the assumption that the system may be described by a stochastic linear system. This assumption allows to get a simple closed form of the rate of {information transfer based on covariances between the observables and their temporal derivatives. We will first test this approach to check the type of information one can gather, and then extend the methodology by using nonlinear observables as in \cite{Pires2024} and \cite{Docquier2024}. We will show that one can indeed have access to the different terms of the entropy budget, provided that the Gaussian assumption is reasonably met.  

\subsection{Liang's approach (linear estimator)}
\label{Linestimation}

Figure~\ref{fig:transl-123456} displays the estimated RET based on the linear assumption for the first 6 variables, $\psi_A, \theta_A, \psi_K,$ $ \theta_K, \psi_L, \theta_L$. The length of the time series used is 500,000 data points sampled every 0.05 time units (equivalent to 0.0055 days), covering 2,750 days. An error bar is also provided through a bootstrap method with replacement (100 bootstrap samples were generated). A first general remark is the absence of information transfer from the second group of variables to the first group. This result is indeed reasonable when one looks at the single RETs from the second group to the first one in Table~\ref{tab:transfer-lin}, which are relatively small. The amplitude of the RETs between the variables of the first group, $\psi_A, \theta_A, \psi_K, \theta_K, \psi_L, \theta_L$, are much larger. The RETs found via the Liang's approach (Figure~\ref{fig:transl-123456}) are however not always of the right amplitude and/or sign compared to the nonlinear case (Table~\ref{tab:transfer-lin}). For the case of $\psi_A$ as a target variable, the values of the linear influence of $\theta_L$ and $\psi_L$ are a little too small in amplitude as compared to the single RETs displayed in Table~\ref{tab:transfer-lin}, even if the equation related to $\psi_A$ is linear: $T(\theta_L \rightarrow \psi_A) = -1.39 10^{-3}$ with the linear approach, while it is equal to $-5.93 10^{-3}$ when including nonlinearities as computed in Section \ref{2.1} (Table~\ref{tab:transfer-lin}), and $T(\psi_L \rightarrow \psi_A) = 6.56 10^{-3}$ with the Liang's method, while it is equal to $1.12 10^{-2}$ with nonlinearities (Table~\ref{tab:transfer-lin}). A similar picture is found when nonlinearities play a role as illustrated for instance with the variable $\psi_K$, where strong dependencies on $\theta_K$ and $\theta_L$ are found, which are quite different to the linear RETs of Table \ref{tab:transfer-lin}.

Figure~\ref{fig:transl-789101112} displays the RETs from the Liang's approach for the second group of variables $\psi_C, \theta_C, \psi_M, \theta_M, \psi_N, \theta_N$. The first group of variables does not influence much the second group. Within the second group, although most of the important variables influencing the target variables are picked up by the Liang's approach, the amplitudes and/or signs are in general not the appropriate ones. 

\begin{figure*}
    \centering
    \includegraphics{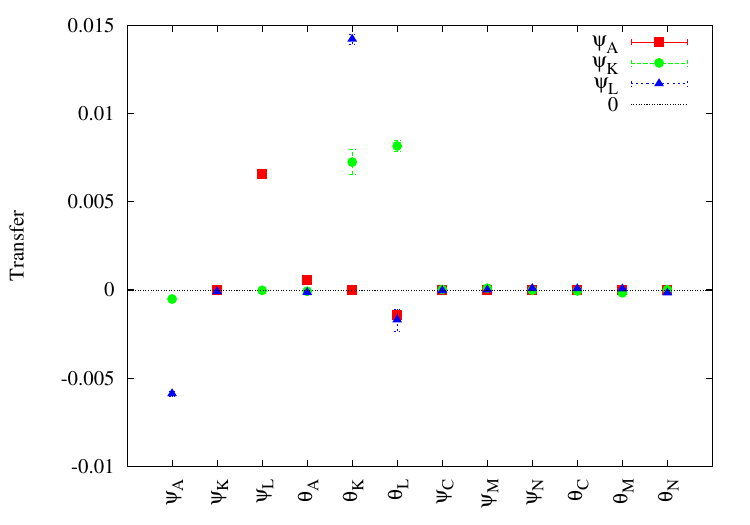}
    \includegraphics{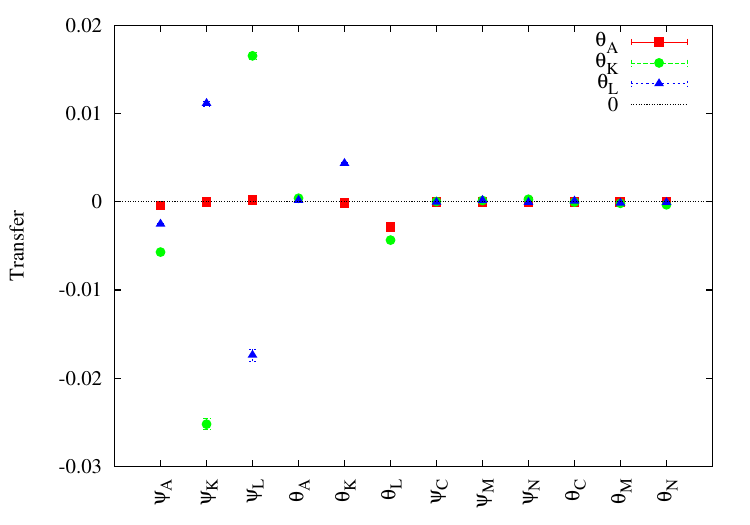}
    \caption{Rate of information transfer from the 12 variables to the target variables $\psi_A, \psi_K, \psi_L$ (top panel), and $\theta_A, \theta_K, \theta_L$ (bottom panel), as obtained with the linear stochastic approximation.}
    \label{fig:transl-123456}
\end{figure*}

\begin{figure*}
    \centering
    \includegraphics{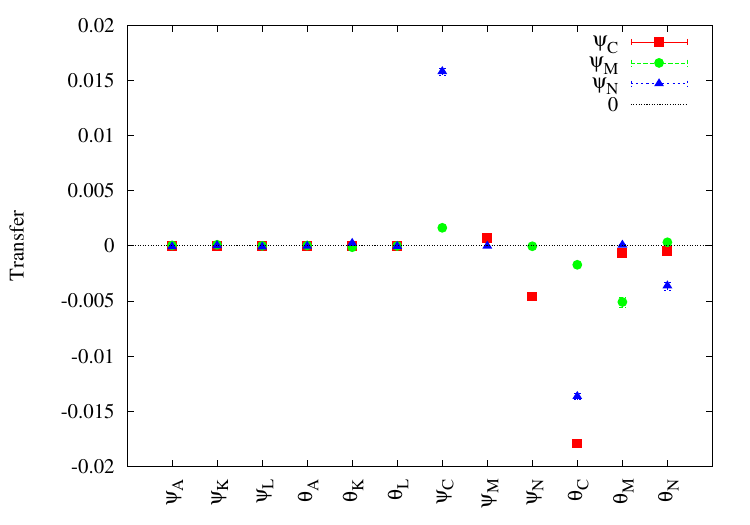}
    \includegraphics{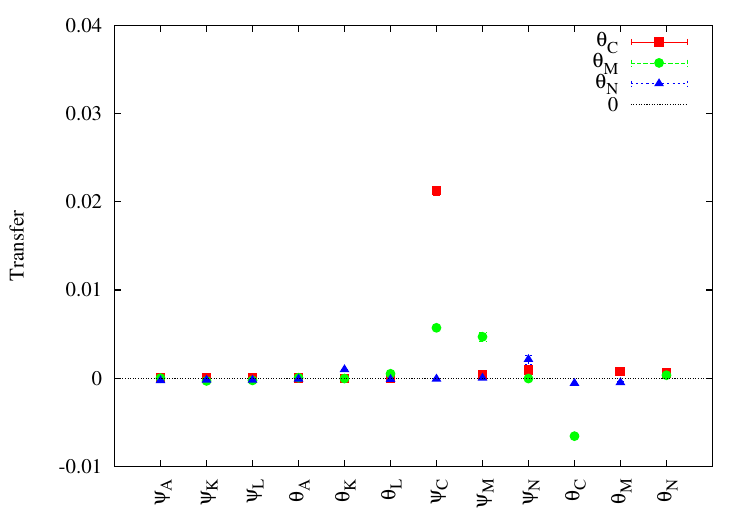}
        \caption{Rate of information transfer from the 12 variables to the target variables $\psi_C, \psi_M, \psi_N$ (top panel), and $\theta_C, \theta_M, \theta_N$ (bottom panel), as obtained with the linear stochastic approximation.}
    \label{fig:transl-789101112}
\end{figure*}

In view of the above considerations on the linear approximation, one may wonder how things may be improved, in particular concerning the amplitude and sign of the rates of information transfer for nonlinear models. A possibility is to fit a nonlinear function to the conditional averages, as done in \cite{Pires2024}, and to compute the RETs. This approach, although successful, may however be strongly dependent on the quality of the fits defined by trial and error, and we will not proceed in that direction here as it was already discussed in \cite{Pires2024}. Another possibility is to keep the linear assumption and to add new observables that take into account the nonlinearities of the dynamical equations. This is the approach adopted in the next section. 

\subsection{Extension of the Liang's approach to the nonlinear terms}
\label{Nonlinestimation}

In the context of the CS model, there are 12 variables in total. If we want to define new nonlinear observables, we could of course take polynomial functions of these 12 variables. If we consider quadratic terms only, it would lead to 78 additional observables when combined blindly. One may also consider higher order polynomial terms of cubic or quartic types. In such cases, the number of observables would grow considerably. It is however obvious that for dynamical systems like the one discussed here, the advection is a dominant nonlinearity, implying that the quadratic terms should play a major role in the dynamics, and the entropy changes. We will therefore consider these quadratic terms only. Finally, as we know the dynamical equations and for the purpose of demonstration, we will define the additional observables as the nonlinearities displayed in Table \ref{tab:entropy-nl}.

In total, we use the 12 original variables plus the 32 nonlinear terms defined in Table \ref{tab:entropy-nl} after removing the coefficients. For instance, one nonlinearity that will be used and which appears in the evolution of $\theta_A$ is $\theta_K \psi_L - \theta_L \psi_K$. The other nonlinearities follow using the same procedure. The 44 observables are now directly used in the approach of \cite{Liang2021}. 

\begin{figure*}
    \centering
    \includegraphics{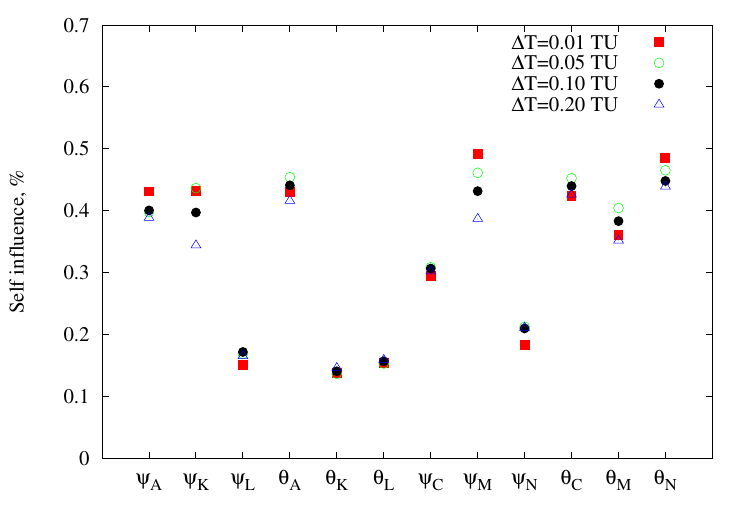}
    \includegraphics{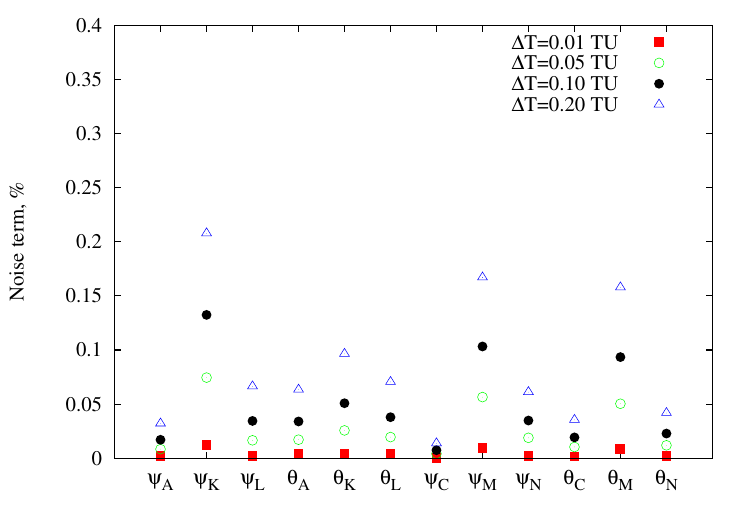}
    \caption{SEG (top panel) and noise amplitude (bottom panel) for the different variables as obtained for four different sampling rates $\Delta t = 0.01, 0.05, 0.10$ and $0.20$ time units. 100,000 data points are used for the estimations.}
    \label{fig:noise-self}
\end{figure*}

As our system is deterministic and all the observables (linear and nonlinear) of the original model are used, the fitting with the linear stochastic model should produce very small stochastic noise, if any. Of course, the way the series is sampled should also have an effect: when the sampling rate is reduced, some information is lost between the successive observation points and the estimation of the time derivatives is less accurate. This should lead to an increase of the noise component. To check this effect, we construct several series sampled at different rates and we apply the linear stochastic model to these series. Four series are built at a sampling rate of $\delta t= 0.01, 0.05, 0.1$ and $0.2$ time unit. For this analysis, 500,000 data points were used for each series. 

Figure \ref{fig:noise-self} (lower panel) displays the percentage contribution of the noise as estimated by fitting the linear stochastic model for the different target variables. As expected, the smaller the $\delta t$, the lower the contribution of the noise. Another interesting result concerns the self influences as obtained for the different sampling times: they are usually all centered around a similar value for the different variables, indicating the robustness of the estimation (Fig.~\ref{fig:noise-self}, upper panel). Note that for the largest sampling rate of 0.20 time units, the estimates of the self-contributions usually differ more from the others. This feature suggests that in order to get robust results in the current context the sampling rate should be relatively high. Thus, in the following, we will use a series sampled every $\delta t= 0.05$ time units in order to keep the estimated noise below 5$\%$. Moreover, as we will use a series of 500,000 points, i.e. 25,000 time units, the series covers 2,750 days, a sufficiently long period to get robust results. 

Figure \ref{fig:transnl-123} displays the RETs for the three target variables $\psi_A, \psi_K, \psi_L$ as a function of the observables used in the adjustment of the linear stochastic model. The 14 first nonlinearities are displayed for visibility. The other ones associated with the dynamics of the second group of variables are not displayed as their contributions are very close to 0. Note that the values cannot be directly compared with the values in Table \ref{tab:transfer-lin} as this table contains RETs that are based on the impact of the causal variables coming from all the terms in the right-hand side of the dynamical equations, or in other words also from the nonlinear terms. One must therefore directly compare with the terms of the entropy balance in Tables \ref{tab:entropy-lin} and \ref{tab:entropy-nl}. 

For the variable $\psi_A$ (Fig.~\ref{fig:transnl-123}), a similar result is found as when using the Liang's approach without nonlinear variables (Fig.~\ref{fig:transl-123456}, upper panel). In this case, the introduction of the nonlinear terms do not improve the quality of the estimations as could be expected since there is no nonlinear terms in its evolution. For the second variable $\psi_K$, already discussed in Section \ref{Linestimation}, the dependencies on $\theta_K$ and $\theta_L$ are now reduced as compared with the results of Fig. \ref{fig:transl-123456}a, getting an appropriate amplitude as given in Table \ref{tab:entropy-lin}. More interestingly, the amplitude of the contributions to the nonlinear terms are captured and are close to the values of the contributions of the nonlinear terms to the entropy budget listed in Table \ref{tab:entropy-nl}. In this case, the signs are also well respected. This discussion can be extended to the other variables as well for this group (see also Fig. \ref{fig:transnl-456}).

\begin{figure*}
    \centering
    \includegraphics{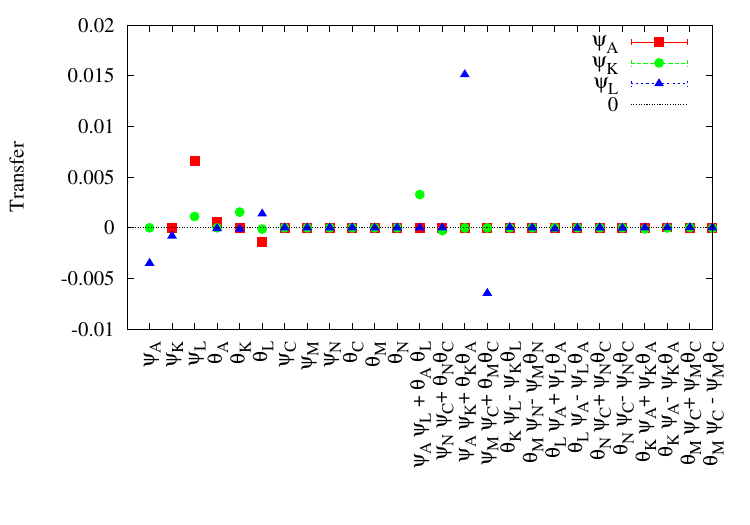}   
    \caption{Rate of information transfer for variables $\psi_A$ (red squares), $\psi_K$ (green full circles) and $\psi_L$ (blue triangles) as obtained with the Liang's approach using the 12 original variables and the 32 nonlinearities appearing in the equations of the model (see Table \ref{tab:entropy-nl} or Appendix A). Only the influence of 16 nonlinearities are displayed as the others show negligible influences. }
    \label{fig:transnl-123}
\end{figure*}

\begin{figure*}
    \centering
    \includegraphics{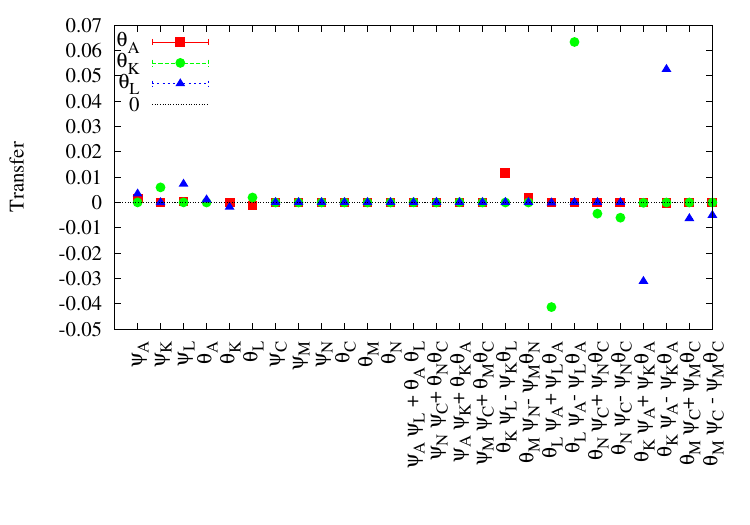}
    \caption{As in Fig.\ref{fig:transnl-123} but for $\theta_A$ (red squares), $\theta_K$ (green full circles) and $\theta_L$ (blue triangles).}
    \label{fig:transnl-456}
\end{figure*}

Now, let us look at the second group. In Figs. \ref{fig:transnl-789} and \ref{fig:transnl-101112}, the RETs are plotted as a function of the original set of variables and the nonlinear terms affecting only this second group. The influence of the nonlinear terms of the first group are close to 0, and are not displayed to get a better visibility. Now the amplitudes and signs are well represented. To give an example, let us consider the variable $\psi_C$, for which the approach used in Section \ref{Linestimation} indicates a dependence on $\theta_C$ close to $-0.02$ (Fig. \ref{fig:transl-789101112}). With the approach using the nonlinear terms, one gets a positive value (see Fig. \ref{fig:transnl-789}), close to the value of the impact of $\theta_C$ reported in Table \ref{tab:entropy-lin}. Furthermore, the contributions of the nonlinear terms to the entropy budget are now accessible with an appropriate amplitude. 

The approach proposed here is very encouraging as it allows to have access to the different contributions of the entropy budget. Still, some differences are present between the real values reported in the Tables \ref{tab:entropy-lin}-\ref{tab:entropy-nl} above and the estimations based on this extended set of variables. The most interesting example is the case of $\psi_A$, for which both methods proposed in this section are not giving the right amplitude (although the signs are correct). To understand this feature, one must go back to the properties of the definition of the RETs given in Eq. (\ref{singleret}), applied to the target variable $\psi_A$. Let us consider the impact of the causal variable $\psi_L$. The RETs in this case is equal to the expectation of the $\psi_A$-derivative of the conditional expectation of $0.5 h_{01} \psi_L$, given $\psi_A$. Figure \ref{fig:psiA-expectation}  displays the conditional expectation as a function of $\psi_A$, which indeed displays a complex dependence, far from the linear dependence we would expect in a purely linear stochastic system. Furthermore, the probability density of $\psi_A$, also shown on this figure, is multimodal. These features leads to an estimate of the RET, far more complicate than a simple linear dependence of the conditional expectation and a Gaussian distribution of $\psi_A$ as assumed by the linear stochastic representation. In such a situation, one must rather rely on a proper estimate of the conditional expectation and the probability density of the target variable.

\begin{figure*}
    \centering
    \includegraphics{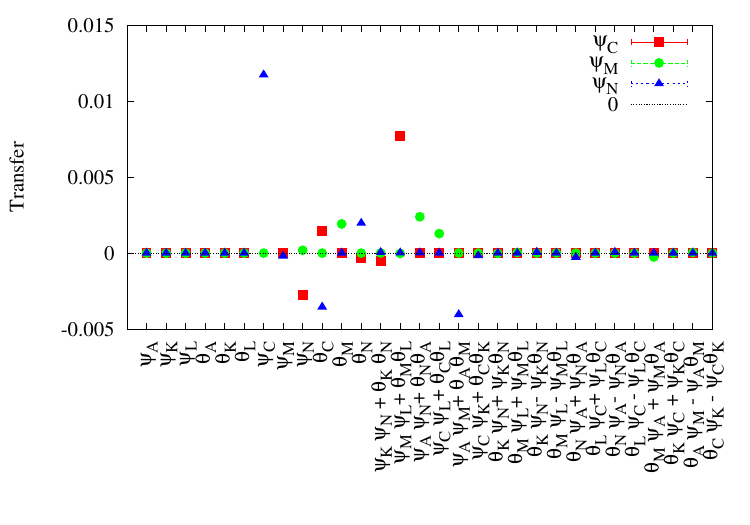}
     \caption{As in Fig.\ref{fig:transnl-123} but for $\psi_C$ (red squares), $\psi_M$ (green full circles) and $\psi_N$ (blue triangles). Note that the nonlinearities displayed on the x-axis are different than in Fig.\ref{fig:transnl-123}.}
    \label{fig:transnl-789}
\end{figure*}

\begin{figure*}
    \centering
    \includegraphics{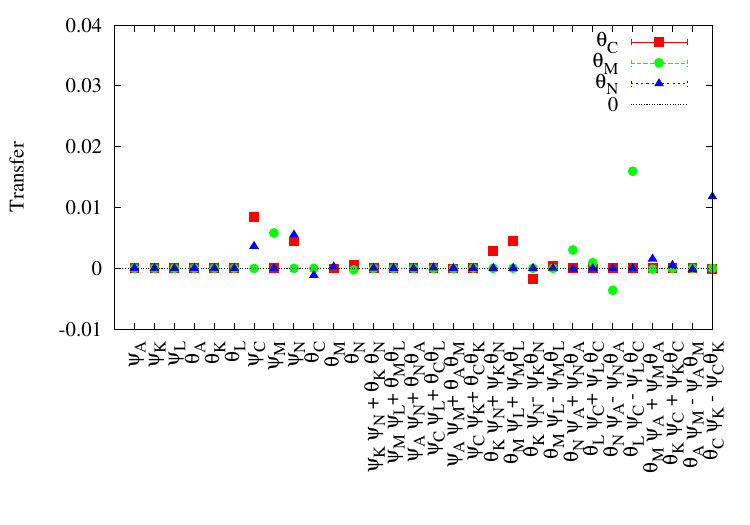}
         \caption{As in Fig.\ref{fig:transnl-789} but for $\theta_C$ (red squares), $\theta_M$ (green full circles) and $\theta_N$ (blue triangles). Note that the nonlinearities displayed on the x-axis are different than in Fig.~\ref{fig:transnl-456}.}
    \label{fig:transnl-101112}
\end{figure*}

\begin{figure*}
    \centering
    \includegraphics{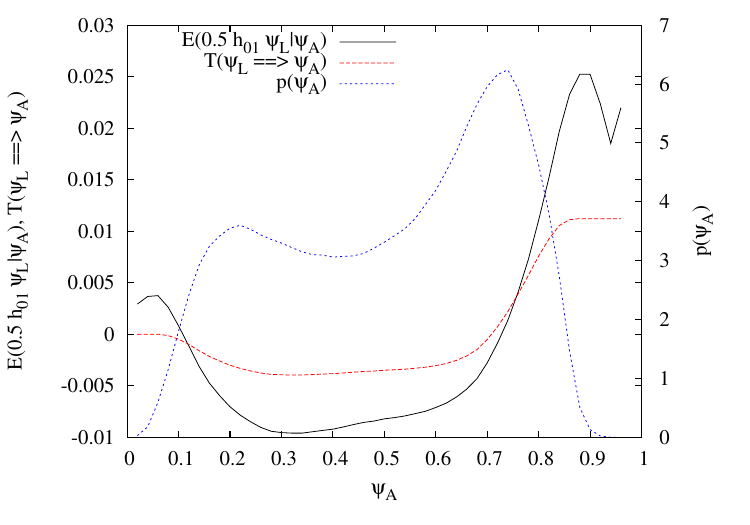}
    \caption{Estimation of the conditional expectation of $ 0.5 h_{01} \psi_L$, given $\psi_A$ (black continuous line); the RET, $T(\psi_L \rightarrow \psi_A)$ (red dotted line); and the probability density of $\psi_A$ (blue dotted line).}
    \label{fig:psiA-expectation}
\end{figure*}

\section{Conclusion}
\label{conclusion}

The information entropy budget and the rate of information transfer between variables has been studied in the context of a reduced-order atmospheric model in which the key ingredients of the dynamics are present, namely the baroclinic and barotropic instabilities, the instability related to the presence of an orography, the dissipation related to the surface friction, and the large-scale forcing inducing a meridional imbalance of energy \citep{Charney1980}. For the parameters chosen in the present work, the solutions display a chaotic dynamics and sensitivity to initial conditions reminiscent of the large-scale atmospheric dynamics in the extra-tropics. The approach used to compute the different terms of the entropy budget and the rate of entropy transfer is based on the computation of conditional expectations, following the recent theory developed in \cite{Pires2024}. 

The detailed information entropy budget analysis of this system reveals that:
\begin{itemize}
    \item The linear rotation terms plays a minor role in the generation of uncertainties as compared to the orography and the surface friction.
    \item By far, the dominant contribution comes from the nonlinear advection terms. The decomposition of the contributions in synergetic and single (impact of each single variable on the target one) rates of information transfer reveals that for some variables the co-variability within the nonlinear terms measured by the synergy dominates, but the opposite may occur.
    \item In view of the importance of the synergetic contributions in the transfer of information/uncertainties, the causality of a variable measured only by single contributions $T(X_j \rightarrow X_i)$ could be misleading or restrictive.
    \item The variables of the model can be split into two groups having little transfer of information from one another: the large-scale group (wavenumber 1 in latitude) and the small(er) scale group (wavenumber 2 in latitude). There are however important synergetic contributions from one group to the other that couple the two groups.
\end{itemize}

The rates of information transfer have been computed based on the knowledge of the dynamical equations of the system. Although important conclusions can be drawn on the role of the different components of the system as detailed above, a natural question is to know whether such type of influences can be extracted from observations. An analysis has been done here using the approach developed by \cite{Liang2021}, which consists in fitting a linear stochastic model to the system, allowing for computing transfer of information between single variables. In the current system, this approach applied to the 12 variables spots relatively well the causal influences, but often does not lead to reasonable  amplitudes and signs of the rate of information transfer. Thus, caution is needed when applying the linear approach designed by \cite{Liang2021} to data displaying high nonlinearities, as already pinpointed in \cite{Pires2024}. 

An extension has been proposed and tested, in which the nonlinear advection terms are also used as potential causal variables, leading to considerable improvements in amplitudes and signs. It allows to compute more appropriately the contributions of the nonlinear terms to the entropy budget. This approach is very promising and should be used in future studies in contexts in which nonlinear contributions are expected. 

Still, the extension of the Liang's approach to nonlinear terms does not improve all rate of information transfer, in particular to strongly non-gaussian variables. Also, the approach does not allow to extract separately single and synergetic contributions. In this context, one needs to rely on the approach proposed by \cite{Pires2024}, in which the conditional expectations should be fitted using nonlinear functions, after which Eqs. (\ref{totalret}-\ref{singleret}) may be used. Further research in that direction are worth performing.

Causal analysis may be used for developing models that will allow for characterizing in a simple way the evolution of the system under scrutiny. An interesting example of that development has been put forward by \cite{Liang2023} in which the causality analysis based on the linear stochastic modeling is used to feed a neural network. This approach is very promising, and the straightforward extension to nonlinear variables discussed in the current analysis could be an easy way to uncover more appropriate causal links.

Another important extension of the current work is to investigate multi-scale systems as for instance land-atmosphere, ocean-atmosphere or cryosphere-atmosphere systems to clarify the importance of such interactions in controlling the atmospheric dynamics. This question will be taken up in the future by first investigating reduced-order coupled models available on the qgs platform (\url{https://github.com/Climdyn/qgs}; accessed 23/03/2024) described in \cite{Demaeyer2020b} for which the dynamical properties originating from coupling are well understood. More complex models fitted to the real Earth System could then be analyzed to clarify the flow of information.  

\section{acknowledgements}
The authors are partly supported by ROADMAP (Role of ocean dynamics and Ocean-Atmosphere interactions in Driving cliMAte variations and future Projections of impact-relevant extreme events; \url{https://jpi-climate.eu/project/roadmap/}), a coordinated JPI-Climate/JPI-Oceans project. David Docquier and Stéphane Vannitsem are also partly supported by the Belgian Federal Science Policy Office under contract B2/20E/P1/ROADMAP. David Docquier is also funded by the Belgian Science Policy Office (BELSPO) under the RESIST project (contract no. RT/23/RESIST). Carlos Pires is supported by Portuguese funds: Fundação para a Ciência e a Tecnologia (FCT) I.P./MCTES through national funds (PIDDAC) – UIDB/50019/2020 (\url{https://doi.org/10.54499/UIDB/50019/2020}), UIDP/50019/2020 (\url{https://doi.org/10.54499/UIDP/50019/2020}) and LA/P/0068/2020 (\url{https://doi.org/10.54499/LA/P/0068/2020}), and the project JPIOCEANS/0001/2019 (ROADMAP).

\newpage
\appendix
\numberwithin{equation}{section}

\section{Dynamical Equations}

\begin{eqnarray}
\frac{d \psi_A}{dt} & = &  -k (\psi_A-\theta_A) + \frac{1}{2} h_{01} (\psi_L-\theta_L)  \label{a1} \\
\frac{d \psi_K}{dt} & = &  - \beta" \alpha (\psi_L \psi_A + \theta_L \theta_A) - \delta \alpha"  (\psi_N \psi_C + \theta_N \theta_C) - k (\psi_K - \theta_K) + \beta_1 \psi_L  \label{a2}\\
\frac{d \psi_L}{dt} & = &    \beta" \alpha (\psi_K \psi_A + \theta_K \theta_A) + \delta \alpha"  (\psi_M \psi_C + \theta_M \theta_C) - k (\psi_L - \theta_L) - \beta_1 \psi_K - \frac{1}{2} h_{n1} (\psi_A - \theta_A)  \label{a3}\\
\left ( 1+\frac{1}{\sigma_0} \right ) \frac{d \theta_A}{dt} & = &  - \frac{\alpha}{\sigma_0} (\theta_K \psi_L - \psi_K \theta_L) - \frac{\alpha'}{\sigma_0} (\theta_M \psi_N - \psi_M \theta_N) + \frac{h"}{\sigma_0} (\theta_A^* - \theta_A) + k \psi_A - (k+2k') \theta_A  \nonumber \\ 
 & & - \frac{1}{2} h_{01} (\psi_L-\theta_L)   \label{a4}\\
\left ( 1+\frac{1-\beta"}{\sigma_0} \right ) \frac{d \theta_K}{dt} & = &  - \beta" \alpha (\theta_L \psi_A + \psi_L \theta_A) - \delta \alpha"  (\theta_N \psi_C + \psi_N \theta_C) + k \psi_K - (k+2k'+\frac{h" (1-\beta")}{\sigma_0}) \theta_K + \beta_1 \theta_L \nonumber \\
& & - \frac{1-\beta"}{\sigma_0} \left [ \alpha (\theta_L \psi_A - \psi_L \theta_A) + \alpha"  (\theta_N \psi_C - \psi_N \theta_C)  \right ] \\
\left ( 1+\frac{1-\beta"}{\sigma_0} \right ) \frac{d \theta_L}{dt} & = & \beta" \alpha (\theta_K \psi_A + \psi_K \theta_A) + \delta \alpha"  (\theta_M \psi_C + \psi_M \theta_C) + k \psi_L - (k+2k'+\frac{h" (1-\beta")}{\sigma_0}) \theta_L - \beta_1 \theta_K \nonumber \\
& & + \frac{1}{2} h_{n1} (\psi_A - \theta_A) - \frac{1-\beta"}{\sigma_0} \left [ \alpha (\theta_A \psi_K - \psi_A \theta_K) + \alpha"  (\theta_C \psi_M - \psi_C \theta_M)  \right ] \\
\frac{d \psi_C}{dt} & = &  \epsilon \alpha" (\psi_K \psi_N + \theta_K \theta_N) - \epsilon \alpha"  (\psi_M \psi_L + \theta_M \theta_L) -k (\psi_C-\theta_C) + \frac{1}{2} h_{02} (\psi_N-\theta_N)   \\
\frac{d \psi_M}{dt} & = &  - \beta' \alpha' (\psi_N \psi_A + \theta_N \theta_A) - \delta' \alpha"  (\psi_L \psi_C + \theta_L \theta_C) - k (\psi_M - \theta_M) + \beta_2 \psi_N  \\
\frac{d \psi_N}{dt} & = &    \beta' \alpha' (\psi_M \psi_A + \theta_M \theta_A) + \delta' \alpha"  (\psi_K \psi_C + \theta_K \theta_C) - k (\psi_N - \theta_N) - \beta_2 \psi_M - \frac{1}{2} h_{n2} (\psi_C - \theta_C) \\
\left ( 1+\frac{1-\epsilon}{\sigma_0} \right ) \frac{d \theta_C}{dt} & = &  \epsilon \alpha" (\theta_K \psi_N + \psi_K \theta_N) - \epsilon \alpha"  (\theta_M \psi_L + \psi_M \theta_L) + k \psi_C - (k+2k'+\frac{h" (1-\epsilon)}{\sigma_0}) \theta_C -  \frac{1}{2} h_{02} (\psi_N-\theta_N) \nonumber \\
& & - \frac{1-\epsilon}{\sigma_0} \left [ \alpha" (\theta_K \psi_N - \psi_K \theta_N) + \alpha"  (\theta_M \psi_L - \psi_M \theta_L)  \right ] \\
\left ( 1+\frac{1-\beta'}{\sigma_0} \right ) \frac{d \theta_M}{dt} & = &  - \beta' \alpha' (\theta_N \psi_A + \psi_N \theta_A) - \delta' \alpha"  (\theta_L \psi_C + \psi_L \theta_C) + k \psi_M - (k+2k'+\frac{h" (1-\beta')}{\sigma_0}) \theta_M + \beta_2 \theta_N \nonumber \\
& & - \frac{1-\beta'}{\sigma_0} \left [ \alpha' (\theta_N \psi_A - \psi_N \theta_A) + \alpha"  (\theta_L \psi_C - \theta_C \psi_L)  \right ] \\
\left ( 1+\frac{1-\beta'}{\sigma_0} \right ) \frac{d \theta_N}{dt} & = &  \beta' \alpha' (\theta_M \psi_A + \psi_M \theta_A) + \delta' \alpha"  (\theta_K \psi_C + \psi_K \theta_C) + k \psi_N - (k+2k'+\frac{h" (1-\beta')}{\sigma_0}) \theta_N - \beta_2 \theta_M  \nonumber \\
& & + \frac{1}{2} h_{n2} (\psi_C-\theta_C) - \frac{1-\beta'}{\sigma_0} \left [ \alpha' (\theta_A \psi_M - \psi_A \theta_M) + \alpha"  (\theta_C \psi_K - \psi_C \theta_K)  \right ]
 \label{a12}
\end{eqnarray}

where the coefficients, 
 \begin{eqnarray}
\beta" & = &  \frac{n^2}{n^2+1}   \\
\beta' & = &  \frac{n^2+3}{n^2+4}    \\
\alpha & = &   \frac{8 \sqrt(2) n}{3 \pi} \\
\alpha' & = &   \frac{32 \sqrt(2) n}{3 \pi} \\
\alpha" & = &   \frac{64 \sqrt(2) n}{3 \pi}\\
\delta & = &   \frac{n^2}{n^2+1}  \\
\delta' & = &   \frac{n^2-3}{n^2+4}  \\
\beta_1 & = & \frac{n}{n^2+1} \frac{L}{R} \frac{cos(\phi_0) }{sin(\phi_0) }  \\
\beta_2 & = &  \frac{n}{n^2+4} \frac{L}{R} \frac{cos(\phi_0) }{sin(\phi_0) }  \\
\epsilon & = & \frac{3}{4} \\
h_{01} & = & \alpha h_k   \\
h_{n1} & = &   \frac{\alpha h_k }{n^2+1}  \\
h_{02} & = &  \frac{\alpha" h_k }{4} \\
h_{n2} & = &  \frac{\alpha" h_k }{n^2+4}  \\
\label{params1}
\end{eqnarray}

and the values of the basic parameters are

\begin{eqnarray}
 2 k & = & C' = k' = h" = 0.0114   \\
\sigma_0 & = &  0.2  \\
h_k & = & 0.1  \\
n & = &  \frac{2 L_y}{L_x} =  1.77  \\
\theta_1^* & = &  0.18 \\
f_0 & = &  0.0001036 s^{-1} \\
L_y & = & 5000 km
\label{params2}
\end{eqnarray}
where $n$ is the aspect ratio, $\sigma_0$ the static stability parameter, $h_k$ the amplitude of the orography, and $f_0$ the Coriolis parameter at $\phi_0=45^\circ$.

\bibliographystyle{wileyqj}
\bibliography{biblio}

\end{document}